\documentclass[aps,pra,twocolumn,showpacs,superscriptaddress,
preprintnumbers,nofootinbib]{revtex4-1}
\pdfoutput=1
\usepackage[T1]{fontenc}
\usepackage{bm,amsmath,amssymb,color,bbold,mathrsfs}
\usepackage[pdftex]{graphicx}
\usepackage[colorlinks=true, pdfstartview=FitV, linkcolor=blue, citecolor=blue, urlcolor=blue]{hyperref}

\DeclareMathOperator\ee{e}

\renewcommand{\Im}{\mathrm{Im}}

\newcommand{\der}{\partial}
\renewcommand{\bar}[1]{\overline{#1}}
\newcommand{\calO}{\mathcal{O}}
\newcommand{\dd}{\mathrm{d}}
\newcommand{\bep}{\begin{pmatrix}} 
\newcommand{\eep}{\end{pmatrix}}

\newcommand{\1}{\mathbb{1}}
\newcommand{\RR}{\mathbb{R}}

\renewcommand{\epsilon}{\varepsilon}

\newcommand{\up}{\uparrow}
\newcommand{\down}{\downarrow}
\newcommand{\muv}{\mu_\textrm{v}}
\newcommand{\mm}{\bar{m}}

\def\ba#1\ea{\begin{align}#1\end{align}}

\def\mkakko#1{\left( #1 \right)}

\renewcommand{\P}{\mathscr{P}}
\newcommand{\A}{\mathcal{A}}

\begin{document}
\preprint{RIKEN-QHP-220}
\title{Mobile impurity in a Fermi sea from the functional 
renormalization group analytically continued to real time}
\author{Kazuhiko Kamikado}
\affiliation{Department of Physics, Tokyo Institute of Technology, Meguro, Tokyo, 152-8551, Japan}
\author{Takuya Kanazawa}
\affiliation{iTHES Research Group and Quantum Hadron Physics Laboratory, RIKEN,
  Wako, Saitama 351-0198, Japan}
\author{Shun Uchino}
\affiliation{RIKEN Center for Emergent Matter Science, Wako,
  Saitama 351-0198, Japan}
\begin{abstract}
 Motivated by experiments with cold atoms, we investigate a mobile
 impurity immersed in a Fermi sea in three dimensions at zero
 temperature by means of the functional renormalization group. We first
 perform the derivative expansion of the effective action to calculate
 the ground state energy and Tan's contact across the polaron-molecule
 transition for several mass imbalances. Next we study quasiparticle
 properties of the impurity by using a real-time method recently
 developed in nuclear physics, which allows one to go beyond the
 derivative expansion. We obtain the spectral function of the polaron,
 the effective mass and quasiparticle weight of attractive and
 repulsive polarons, and clarify how they are affected by mass
 imbalances.
\end{abstract}
\maketitle 

\section{Introduction}
The concept of quasiparticles has become a cornerstone of quantum
many-body physics.  One of the fundamental problems is to understand
properties of a \emph{polaron}, a single mobile impurity in a bath of
majority particles
\cite{Mitra1987,feynman1998,Devreese0904,Grusdt:2015fqa}.  While such an
impurity effectively behaves as a free particle, renormalization effects
due to interactions between the impurity and the bath are known to be
significant.

The polaron problem had been discussed originally in the context of
solid state physics, where an electron interacts with a bath of phonons,
that is, bosons \cite{mahan2000}.  Recently, experiments with ultracold
atoms have allowed us to consider a similar problem with fermions: an
impurity immersed in a bath of fermions \cite{Massignan1309,Lan1404}.
Experiments using a mixture of two different hyperfine states as well as
a hetero-nuclear mixture have been performed in two and three dimensions
\cite{Schirotzek0902,Nascimbene2009, Kohstall1112,Koschorreck1203}.  The
$s$-wave scattering length characterizing the interaction between
impurities and fermions in the bath can be tuned via the Feshbach
resonance (FR) technique. This opens up a way to realize a novel
strongly-coupled impurity system, whose understanding requires a
non-perturbative analysis.

A number of different techniques in quantum many-body physics have been
applied to the analysis of the polaron problem.  A recently developed
powerful numerical scheme, called the diagrammatic Monte Carlo (diagMC)
method, was successfully applied to the broad-FR case
\cite{Prokofiev2008a,Prokofiev2008b,Vlietinck1302,Kroiss1411}.
Analytical methods including a variational calculation and a many-body
$T$-matrix approach were also employed
\cite{Combescot0702,Combescot0804,Massignan:2008zz,
Combescot0907,Mora0908,Punk0908,CuiZhai1001,
Bruun1003,Massignan1102,Mathy1002,Trefzger1112}.

The functional renormalization group (FRG) based on an exact flow
equation \cite{Wetterich:1992yh,Morris:1993qb,
Berges:2000ew,Delamotte:2007pf,Metzner:2011cw} offers an alternative
route to tackling quantum many-body problems.  Since FRG is built on a
philosophy different from the conventional methods above, approximations
in FRG may provide new insights into the mobile impurity problem.  In
\cite{Schmidt:2011zu}, Schmidt and Enss examined this problem in the
broad-FR limit using a Blaizot--M$\acute{\rm e}$ndez-Galain--Wschebor
(BMW)-type FRG \cite{Blaizot:2005xy} combined with the Pad\'{e}
approximation for analytic continuation to real time \cite{Vidberg1977}.
Their approach yields accurate results at imaginary time, but it comes
at a high computational cost.

In this paper, we study the Fermi polaron problem using FRG based on the
derivative expansion, which is numerically much less expensive than the
BMW approach.  While the derivative expansion of the scale-dependent
effective action is able to capture critical phenomena accurately
\cite{Berges:2000ew,Delamotte:2007pf}, it is inadequate to describe
spectral properties of quasiparticles because it fails to capture higher
excited states and a continuum incoherent background.  Recently,
however, schemes to circumvent such difficulties of the derivative
expansion on the basis of analytic continuation of flow equations to
real time has been developed and tested in nuclear physics
\cite{Kamikado:2012bt,Kamikado:2013sia,Tripolt:2013jra,Tripolt:2014wra}.
In this work, we apply this scheme to the polaron problem at zero
temperature, and reveal not only thermodynamic properties of the system
but also spectral properties of quasiparticles including highly excited
states, thus going far beyond the conventional realm of a derivative
expansion. We remark that this is the first application of this scheme
to a nonrelativistic, experimentally accessible system. This provides a
viable alternative to other RG-based approaches to nonequilibrium
physics \cite{RevModPhys.80.395,Berges201237}.

This paper is organized as follows.  In Section \ref{sc:model} we
present our model Hamiltonian.  In Section \ref{sc:floweq} we explain
elements of FRG and describe flow equations in our formulation.  Section
\ref{sc:results} shows our main results on thermodynamic and spectral
properties of the Fermi polaron system.  Section \ref{sc:disc} is
devoted to a summary.  In Appendixes we discuss technical details of our
formulation.

\section{\label{sc:model}Model}
We consider two-component fermions with an $s$-wave interaction.  Since
the $s$-wave scattering length is experimentally tuned via a molecule
state in the closed channel, the natural model reflecting the
microscopic dynamics of the FR is the so-called two-channel model 
(with $\hbar=1$):
\ba
	\!\! S & =\int_{\mathbf{x},\tau}
	\bigg[
	\sum_{\sigma=\up,\down}
	\psi^{*}_{\sigma}\mkakko{
		\partial_{\tau}-\frac{\nabla^2}{2M_{\sigma}}
		- \mu_{\sigma}} \psi_{\sigma} 
	\notag
	\\
	& \quad 
		+ \phi^{*}\mkakko{
		\partial_{\tau}
		- \frac{\nabla^2}{2M_\phi}
		- \mu_{\phi}
	}\phi + h(\psi^{*}_{\up}
	\psi^{*}_{\down}\phi+\text{H.c.}) \bigg],
	\label{eq:act}
\ea
where $\psi_\up$ is the majority atom forming a non-interacting Fermi
sea, $\psi_\down$ is the impurity atom, $\phi$ is the molecule field
made of $\up$ and $\down$ atoms, and $M_{\phi}=M_{\up}+M_{\down}$.  The
Yukawa (Feshbach) coupling $h$ accounts for the formation and
dissociation of molecules. Physically, $h$ is related to the width of
the FR. It is known \cite{Gurarie0611,Braaten:2007nq} that the large-$h$
limit corresponds to a broad FR where the scattering amplitude is
described solely by the $s$-wave scattering length, while the small-$h$
limit corresponds to a narrow FR where in addition to the $s$-wave
scattering length, the effective range is involved in the scattering
amplitude.
The chemical potential $\mu_\up$ for the majority atoms is
adjusted to the Fermi energy $E_{F}=k^2_{F}/(2M_{\up})$; the
renormalization due to the impurity can be neglected in the
thermodynamic limit.  Determination of $\mu_{\down}$ and $\mu_{\phi}$
will be discussed in subsequent sections.

In what follows, we use natural units $\hbar=1$ and set
\ba
	2M_{\up}=1,~~2M_{\down}=\alpha \quad \text{and} \quad 2M_\phi=1+\alpha\,. 
\ea

\section{\label{sc:floweq}Flow equations in the functional renormalization group}
The starting point of FRG is the following functional differential
equation \cite{Nicoll:1977hi,Wetterich:1992yh,Morris:1993qb}:
\ba
	\partial_k\Gamma_k=\frac{1}{2}
	\text{STr}[(\Gamma^{(2)}_k+R_k)^{-1}\partial_kR_k],
	\label{eq:wetterich}
\ea
where $\Gamma_k$ and $\Gamma^{(2)}_k$ represent the
average action at a scale $k$ and an inverse of the two-point 
function, respectively. 
The symbol STr denotes the supertrace where
the summation over momenta and frequencies, internal indices,
and fields is taken.
Notice that in the supertrace, a minus sign for fermions is necessary.
An important observation is that although the above flow equation
has a one-loop structure, it contains the full field-dependent propagator
and therefore allows one to incorporate nonperturbative effects.

The average action $\Gamma_k$ includes all fluctuations with momenta
$q\gtrsim k$.  At the ultraviolet (UV) scale $k=\Lambda$, $\Gamma_k$
reduces to the classical action, while at the infrared (IR) scale
$k\to0$ one obtains the full quantum effective action.  These properties
of the average action can be ensured through the use of a regulator
function $R_k$, which possesses the following properties:
$\lim_{k\to\infty} R_k(p)=\infty$, $\lim_{k\to0}R_k(p)=0$, and
$\lim_{p\to0} R_k(p)>0$ \cite{Berges:2000ew}.  While in general, the
flow equation itself depends on the choice of the regulator function,%
\footnote{For a detailed discussion on the use of 4d regulators, see \cite{Pawlowski:2015mia}.}
resultant physical properties in the IR limit are expected to be
unaltered.  In this paper, we choose the so-called 3d sharp cutoff
regulator: in the case of bosons, $R_k(p)=0$ for $k\leq|\mathbf{p}|$ and
$R_k(p)=\infty$ for $|\mathbf{p}|<k$. In the case of finite-density
fermions, the cutoff should be imposed on momenta measured from the
Fermi surface \cite{Diehl:2009ma,Braun:2011pp,Boettcher:2012cm}.  This
sharp regulator is particularly useful in treating the flow equations
analytically.

As the exact flow equation \eqref{eq:wetterich} is too complicated to
solve exactly, some truncation procedure is practically required.  We
adopt the following truncation for the imaginary-time average effective
action
\ba
	\Gamma_{k} & =\int_{P}\Big[
	\psi^{*}_{\up}
	(-ip_0+\mathbf{p}^2 - k_F^2) \psi_{\up}+
	\psi^{*}_{\down}\P_{\down,k}(P)
	\psi_{\down}
	\nonumber
	\\
	& \qquad 
	+\phi^{*}\P_{\phi,k}(P)\phi+
	h(\psi^{*}_{\up}
	\psi^{*}_{\down}\phi+\text{H.c.}) 
	\Big]\,,
	\label{eq:average-action}
\ea
with $P=(p_0,\mathbf{p})$.  Here, $\P_{\down,k}$ and $\P_{\phi,k}$ are
inverse propagators of the impurity and the molecule, respectively.  For
the sake of simplicity we omit interactions
$\psi^{*}_{\up}\psi^{*}_{\down}\psi_{\down} \psi_{\up}$ and
$\psi^{*}_{\up}\phi^{*}\psi_{\up}\phi$, and neglect the renormalization
of $h$.

\begin{figure}[t]
	\centering
	\includegraphics[width=80mm]{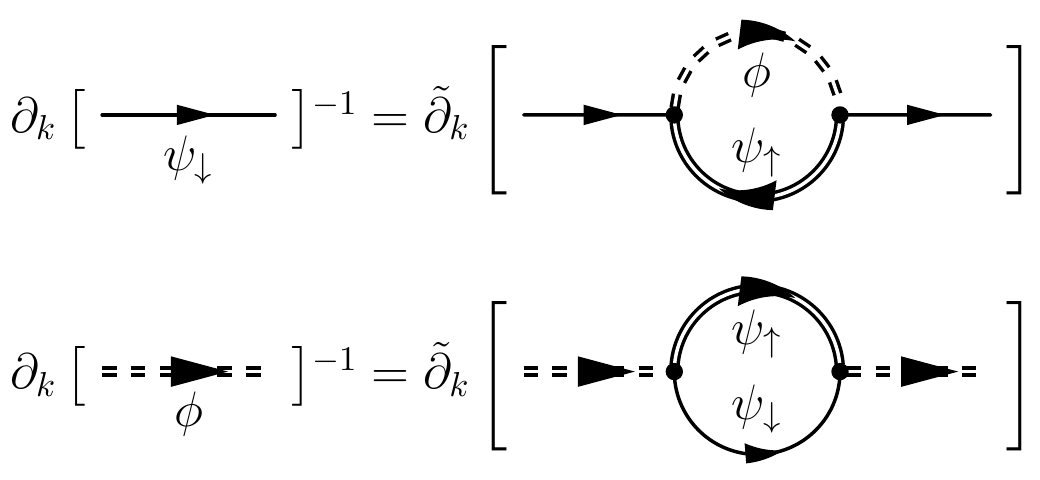}
	\caption{\label{fg:flowdiag}
		Diagrams corresponding to the flow equations \eqref{eq:rfd} of FRG. 
		All the propagators in the loops are regularized with $R_k$. 
	}
\end{figure}

By using the truncation \eqref{eq:average-action} we obtain the
functional flow equations \cite{Schmidt:2011zu}, depicted schematically
in Fig.~\ref{fg:flowdiag}:
\begin{subequations}
 \ba
  \partial_k \P_{\down,k}(P) & =h^2\tilde{\partial}_k\int_Q
  G^c_{\phi,k}(Q)G^c_{\up,k}(Q-P),
  \\
 \partial_k \P_{\phi,k}(P) & = -h^2\tilde{\partial}_k\int_Q
  G^c_{\down,k}(Q)G^c_{\up,k}(P-Q),
  \ea
  \label{eq:rfd}%
\end{subequations}
where $\tilde{\partial}_k$ represents a derivative that only acts on the
$k$-dependence of the regulator $R_k$, and $G^c_k\equiv 1/(\P_k+R_k)$.

Solving \eqref{eq:rfd} is a challenging numerical problem.  Instead of
solving these integro-differential equations by brute force we shall
proceed in two steps as follows \cite{Kamikado:2012bt, Kamikado:2013sia,
Tripolt:2013jra}: in the first step, we solve \eqref{eq:rfd} in the
leading-order derivative expansion. In the second step, we substitute
solutions obtained from the derivative expansion into the RHS of
\eqref{eq:rfd} and integrate both sides from $k=\Lambda$ to $k=0$. In
this way one can find $\P_k$ without assuming any specific Ansatz for
the form of $\P_k$. In principle one may feed the obtained $\P_k$ into
the RHS of \eqref{eq:rfd} and integrate over $k$ again.  If this
procedure is iteratively repeated sufficiently many times, one would
gain exact solutions to \eqref{eq:rfd}. However, as previous analyses
\cite{Kamikado:2012bt} indicate, the first iteration already yields
reasonably accurate results and we explicitly demonstrate this in the
following.

As the first step let us consider a truncation of $\P_k$ in the 
leading-order derivative expansion which respects the Galilean invariance of the theory, 
\begin{subequations}
\ba
	\P_{\down,k}(P) & =A_{\down,k}(-ip_0+\mathbf{p}^2/\alpha) + m_{\down,k}^2\,,
	\\
	\P_{\phi,k}(P) & =A_{\phi,k}[-ip_0+\mathbf{p}^2/(1+\alpha)] + m_{\phi,k}^2\,,
\ea
\end{subequations}
where $A_{\downarrow,k}$ ($A_{\phi,k}$) and $m^2_{\downarrow,k}$ ($m^2_{\phi,k}$)
represent the wave-function renormalization and gap
for the impurity (the molecule), respectively.  \newpage Then the
regulated Green's functions read
\begin{subequations}
	\ba
		G^c_{\up,k}(P) & = \frac{\theta(|\mathbf{p}^2 - k_F^2|-k^2)}
		{-ip_0+\mathbf{p}^2 - k_F^2} \,,
		\\
		G^c_{\down,k} (P) & = \frac{\theta(|\mathbf{p}|-k)}
		{ A_{\down,k}(-ip_0+\mathbf{p}^2/\alpha) + m_{\down,k}^2 }\,,
		\\
		G^c_{\phi,k}(P) & = \frac{\theta(|\mathbf{p}|-k)}
		{ A_{\phi,k}[-ip_0+\mathbf{p}^2/(1+\alpha)] + m_{\phi,k}^2 } \,, 
	\ea
	\label{eq:GGG}%
\end{subequations}
where $\theta(x)$ is the usual Heaviside step function. 

Plugging \eqref{eq:GGG} into \eqref{eq:rfd} we straightforwardly obtain
\begin{widetext}
\begin{subequations}
	\label{eq:fl0}
	\ba
		\der_k \P_{\down,k}(P) 
		& = \frac{h^2k}{4\pi^2 A_{\phi,k}} \Bigg\{
		k \int_{-1}^1\dd x 
		\frac{
			\theta(k_F^2 - 2k^2 - p^2 + 2pkx)
		}{
			- ip_0 + k_F^2 - p^2 + 2pkx  
			- \frac{\alpha}{1+\alpha}k^2 + m_{\phi,k}^2/A_{\phi,k}
		}
		\notag
		\\
		& \quad + \sqrt{k_F^2-k^2}\,
		\theta(k_F^2-k^2) \int_{-1}^{1} \dd x  
		\frac{
			\theta\big(k_F^2 - 2k^2+p^2+2px \sqrt{k_F^2-k^2}\,\big)
		}{
		  - ip_0 + \frac{p^2+\alpha k^2+k_F^2+ 2px\sqrt{k_F^2
	              -k^2}}{1+\alpha} 
			 + m_{\phi,k}^2/A_{\phi,k}
		}\Bigg\}
		\,, 
		\\
		\der_k \P_{\phi,k}(P) 
		& = \frac{h^2 k}{4\pi^2 A_{\down,k}}\Bigg\{
		k \int_{-1}^{1}\dd x 
		\frac{
			\theta(p^2 - 2pkx - k_F^2)
		}{
			-ip_0 + p^2 -2pkx - k_F^2 + \frac{1+\alpha}{\alpha} k^2 
			+ m_{\down,k}^2/A_{\down,k}
		}
		\notag
		\\
		& \quad + \sqrt{k_F^2+k^2}
		\int_{-1}^{1}\dd x 
		\frac{
			\theta\big(  k_F^2 + p^2 + 2px\sqrt{k_F^2+k^2} \, \big)
		}{
		  -ip_0
	          + \frac{k_F^2 + (1+\alpha)k^2 + p^2 + 2px\sqrt{k_F^2+k^2}}{\alpha}
			+ m_{\down,k}^2/A_{\down,k}
		}
		\Bigg\}\,.
	\ea
\end{subequations}
\end{widetext}
To speed up numerical analysis we perform 
the remaining integral over $x$ analytically with the formula
\ba
	\label{eq:3dintformula}
	\int_{-1}^{1}\dd x\frac{\theta(a_1 + a_2 x)}{a_3 + a_2x}
	=\Gamma(a_1, a_2, a_3) 
\ea
where $a_1,a_2\in\RR$\,,  $a_2\ne 0$, $\Im\ a_3\ne 0$\,, and 
\begin{multline}
	\Gamma (a_1, a_2, a_3) \equiv
	\frac{\theta(a_1+|a_2|)}{|a_2|}\Big\{ \log(|a_2|+a_3) 
	\\
	- \log\big[\text{max}(0, a_1-|a_2|)+ a_3 - a_1\big]
	\Big\} \,.
\end{multline}
The resulting form of \eqref{eq:fl0} is presented in Appendix \ref{ap:2pfloweq}. 

In units where $\hbar=2M_{\up}=1$, all dimensionful quantities can be
measured in units of the Fermi momentum $k_F$ of the $\up$ atoms.  It is
then useful to define
\ba
	\begin{split}
		& ~~~  t \equiv \log \frac{k}{k_F}\,, 
		\quad 
		\der_t = k \frac{\der}{\der k}\,,
		\quad 
		\hat{h} \equiv \frac{h}{\sqrt{k_F}}\,,
		\\ 
		& \qquad ~
		\hat m_{\down,k} \equiv \frac{m_{\down,k}}{k_F}\,, 
		\quad
		\hat m_{\phi,k} \equiv \frac{m_{\phi,k}}{k_F}\,.
	\end{split}
\ea
The flow equations for $m_{\down,k}$, $m_{\phi,k}$, $A_{\down,k}$ and
$A_{\phi,k}$ can be derived via expansion of \eqref{eq:fl0} around the
low-frequency and low-momentum limit. The results read
\begin{subequations}
	\label{eq:DEfl}
	\ba
		\der_t A_\down &= - 
		\theta(1-2\ee^{2t})
		\frac{\hat h^2 \ee^{2t}}{2\pi^2 A_\phi}
		\Bigg[
		\frac{\ee^{t}}{
			(1-\frac{\alpha}{1+\alpha}\ee^{2t}
			+\hat m_\phi^2/A_\phi)^2
		}   
		\notag
		\\
		& \qquad +
		\frac{\sqrt{1-\ee^{2t}}}{
			(\frac{1+\alpha\ee^{2t}}{1+\alpha}
			+\hat m_\phi^2/A_\phi)^2
		} 
		\Bigg],
		\\
		\der_t \hat m_\down^2 & = 
		\theta(1 - 2\ee^{2t})
		\frac{\hat h^2 \ee^{2t}}{2\pi^2 A_\phi}
		\Bigg[
			\frac{\ee^t}{1-\frac{\alpha}{1+\alpha}\ee^{2t}
			+\hat m_\phi^2/A_\phi}
		\notag
		\\
		& \qquad + 
		  	\frac{\sqrt{1-\ee^{2t}}}{
		  		\frac{1+\alpha \ee^{2t}}{1+\alpha}
				+\hat m_\phi^2/A_\phi
			}
		\Bigg],
		\\
		\der_t A_\phi & = - \frac{\hat h^2 \ee^{2t}}{2\pi^2A_\down}
		\frac{\sqrt{1+\ee^{2t}}}{
			\big[ \frac{1+(1+\alpha)\ee^{2t}}{\alpha}
			+\hat m_\down^2/A_\down \big]^2
		} \,,
		\\
		\der_t \hat m_\phi^2 & = 
		\frac{\hat h^2\ee^{2t}}{2\pi^2 A_\down}
		\frac{\sqrt{1+\ee^{2t}}}{
			\frac{1+(1+\alpha)\ee^{2t}}{\alpha}
	          + \hat m_\down^2/A_\down
		} \,.
		\label{eq:mpfll}
	\ea
\end{subequations}
For $\alpha=1$, \eqref{eq:DEfl} reduce to 
\cite[eq.(A2)]{Schmidt:2011zu}. 

The initial conditions at the scale $k=\Lambda$ are given by
\begin{gather}
	A_{\down,\Lambda}=A_{\phi,\Lambda}=1\,,
	\label{pp1}
	\\
	\hat m_{\down,\Lambda}^2 = - \hat\mu_\down \,, 
	\label{pp3}
	\\  
	\hat m_{\phi,\Lambda}^2 = \frac{\alpha}{4\pi(1+\alpha)} \hat h^2
	\mkakko{\frac{2}{\pi}\ee^{t_0}-\frac{1}{k_F a}} 
	- 1 - \hat\mu_\down ,
	\label{pp4}
\end{gather}
where $\displaystyle t_0 \equiv \log (\Lambda/k_F)$.  These initial
conditions are derived in Appendix \ref{ap:vaclim} for completeness.
The value of $\hat\mu_\down$ should be chosen so that $\text{Min}(\hat
m^2_{\down,k=0},\hat m^2_{\phi,k=0}) =0$ \cite{Schmidt:2011zu}; see
Sec.~\ref{sc:results} for a more precise explanation.

Once the solutions to \eqref{eq:DEfl} are obtained for all $k$, we
substitute them into the RHS of \eqref{eq:fl0} and then perform analytic
continuation of \eqref{eq:fl0} to real frequency $\omega$ via
\ba
	\label{eq:ancont}
	p_0 = - i(\omega+i\epsilon)
\ea
where $\epsilon>0$ is an infinitesimal constant (see
Appendix~\ref{ap:2pfloweq} for flow equations at real frequency).
Afterwards we integrate both sides of \eqref{eq:fl0} from $k=\Lambda$
down to $k=0$ and finally obtain the retarded two-point functions
$\hat{\P}^{\rm R}_{\down/\phi} (\hat{\omega},\hat{p})=\P^{\rm R}_{\down/\phi} (\hat{\omega},\hat{p})/k_F^2$ in Minkowski
spacetime, with $\hat\omega\equiv\omega/k_F^2$ and $\hat{p}\equiv
p/k_F$.  The spectral functions are obtained as
\ba
	\quad \A_{\down/\phi}(\hat\omega,\hat p) 
	& = \frac{1}{\pi}\Im \frac{1}{
		\hat{\P}^{\rm R}_{\down/\phi}(\hat{\omega},\hat{p})
	}\,.
\ea
They should not be confused with the wave function renormalization
$A_{\down/\phi}$.

In this method, in which the analytic continuation is done directly at
the level of the flow equation, one can entirely avoid a numerically
expensive implementation of approximate analytic continuation of
Euclidean correlators to real time, such as the Pad\'e approximation
\cite{Vidberg1977} and the maximum entropy method
\cite{Jarrell:1996rrw}. In addition, we note that the computional time in this method
is remarkably fast. This is due to the fact that
our flow equations consist only of  $A_{\downarrow/\phi,k}$ and $m_{\downarrow/\phi,k}$,
each of which can be determined with the derivative expansion, and we can analytically perform the
integrals in momentum and frequency with Eq.~\eqref{eq:3dintformula}.
This is in contrast to the approach in  Ref.~\cite{Schmidt:2011zu}, where
one must take every points in $(\omega,p)$ plane and perform the loop integral in flow equations numerically.
These are the major advantages of the present
scheme.

\section{\label{sc:results}Results}
In this section we proceed to numerical results.  Parameters in the
present model have been chosen as follows.
\begin{itemize}
 \item The (normalized) inverse scattering length $(k_Fa)^{-1}$ is
       varied from negative to positive values across the unitarity
       limit.  This enables us to probe the polaron-molecule transition.
 \item As for the mass ratio $\alpha = M_{\down}/M_{\up}$, we will
       examine three cases, $\alpha = 6.6,\; 1 $ and $0.5$ covering both
       heavy and light impurities.  In choosing $\alpha = 6.6\simeq
       40/6$ we have in mind $^{\rm 40}$K atoms immersed in the Fermi
       sea of $^6$Li atoms \cite{Kohstall1112}.%
       \footnote{We note that
       the experimental system of \cite{Kohstall1112} corresponds to a
       narrow FR, and also at nonzero temperature. Thus a direct
       comparison of \cite{Kohstall1112} with our numerical results is
       not possible.}
 \item In this work we limit ourselves to a broad-FR system, which
       corresponds to the large-$h$ limit of \eqref{eq:act}
       \cite{Gurarie0611,Braaten:2007nq}.  We have confirmed that the
       numerical results become insensitive to $\hat{h}$ provided
       $\hat{h}\gtrsim 150$.  Throughout this section we choose
       $\hat{h}=300$.
 \item We set $\Lambda=10^3k_F$ as in \cite{Schmidt:2011zu}, which
       implies \ba t_0=\log 10^3\simeq 6.907\,.  \ea The flows are
       solved for $t_{\min} \leq t \leq t_0$.  The lower limit
       $t_{\min}$ should be chosen small enough to ensure convergence of
       the flow. In this work, we set $t_{\rm min}=-5.0$. We have
       confirmed that at $t_{\rm min}$, all flowing parameters already
       reach asymptotic values; see
       Appendix~\ref{sec:flowing-parameters} for an explicit check.
 \item We set $\hat{\epsilon}\equiv \epsilon/k^2_F = 10^{-3}$ throughout this work.  For
       completeness, the $\hat{\epsilon}$ dependence of spectral
       functions is examined in Appendix~\ref{sec:epsil-depend-spectr}.
\end{itemize}

 \subsection{Thermodynamic properties}
 We first examine thermodynamic properties of the impurity system.
 \begin{figure}[bt]
	  \begin{center}
	  	\includegraphics[width=0.95\columnwidth]{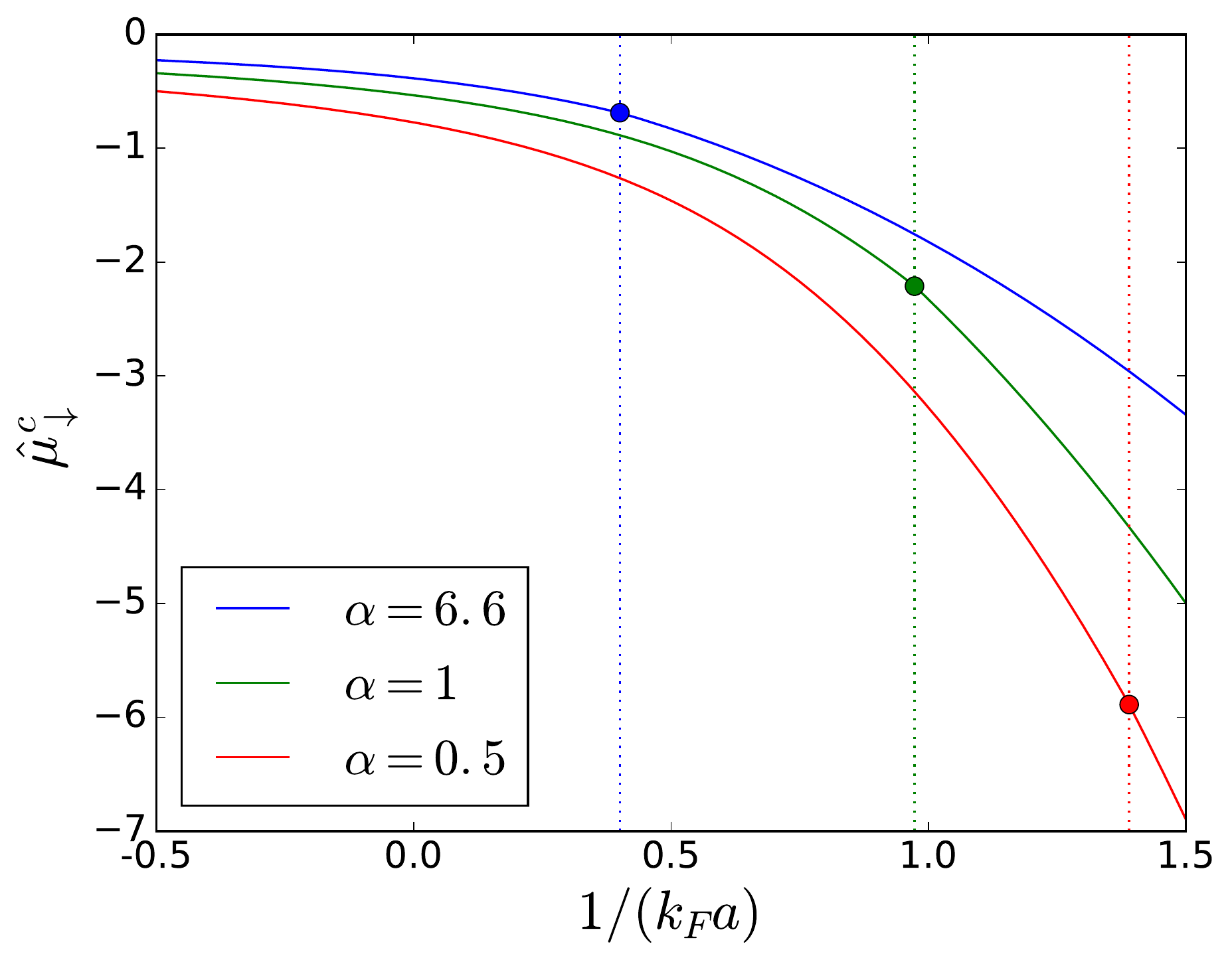}
	  \end{center}
	  \vspace{-1.5\baselineskip}
	  \caption{\label{fig:ground_state_energy} Ground state energy
	    of the impurity system for $\alpha=6.6$ (upper curve),
            $\alpha=1$ (middle curve), and $\alpha=0.5$ (lower curve).
            Vertical dotted lines represent the
	  location of the polaron-molecule transition for each
	  $\alpha$.}
 \end{figure}
 In order to realize a system with a single $\down$ atom in the medium
 of $\up$ atoms, we tune the chemical potential of the $\down$ atom. The
 ground state energy is determined as the \emph{minimal}
 $\hat{\mu}_{\down}^c$ that satisfies the following condition
 \cite{Schmidt:2011zu}
\ba
	\hat{m}^2_{\down,k=0} \ge 0~~\text{and}~~ \hat{m}^2_{\phi,k=0} \ge 0 
	\quad \; \text{for}~~ \forall \hat{\mu}_{\down} \ge
	\hat{\mu}^c_{\down}\,.
\ea
 If $\hat{m}_{\down,k=0}^2=0$ and $\hat{m}_{\phi,k=0}^2>0$ at
 $\hat{\mu}_\down=\hat{\mu}_{\down}^c$\,, the system is on the polaronic
 side, i.e, forming a molecule is energetically more costly than a
 polaron. On the other hand, if $\hat{m}_{\down,k=0}^2>0$ and
 $\hat{m}_{\phi,k=0}^2=0$ at $\hat{\mu}_{\down}^c$, the system is on the
 molecule side and a molecule is formed in the ground state.  
 
 In Table.~\ref{tab:simulation_points}, we summarize the resulting values
 of the polaron-molecule transition points where both the polaron and
 the molecule are gapless in the IR limit.
 \begin{table}[h]
    \caption{\label{tab:simulation_points}
		   Polaron-molecule transition points for various mass imbalances.}  
	 \begin{center}
		   \begin{tabular}[t]{|c||c|}
			    \hline
			    $\alpha$& Transition point\\
			    \hline
			    \hline
			    6.6  &$(k_Fa)^{-1}=0.4006$ \\
			    \hline
			    1.0  &$(k_Fa)^{-1}=0.973$ \\
			    \hline
			    0.5 &$(k_Fa)^{-1}=1.390$ \\
			    \hline
		   \end{tabular}
	 \end{center}
 \end{table}
 The tendency that the polaron-molecule transition moves to 
 higher $1/(k_Fa)$ for a lighter impurity is consistent with previous
 studies \cite{Combescot0907,Massignan1112,Trefzger1112}.  
 The polaron-molecule transition in our leading-order derivative
 expansion for $\alpha=1$ occurs at $(k_Fa)^{-1}=0.973$, which compares
 well with the prediction of the diagMC, 0.90(2)
 \cite{Prokofiev2008a,Prokofiev2008b} and that of the BMW-type FRG, 0.904(5) \cite{Schmidt:2011zu}.
 Close values are obtained in
 other approaches as well
 \cite{Combescot0907,Mora0908,Punk0908,Massignan1102,Trefzger1112,Vlietinck1302}.

 Figure~\ref{fig:ground_state_energy} displays 
 the ground state energy $\hat{\mu}^{c}_{\down}$ of the impurity 
 $\hat{\mu}^{c}_{\down}$.%
 \footnote{We note that the ground state energy 
 in FRG slightly depends on the choice of the regulator $R_k$ \cite{Pawlowski:2015mlf}.}  
 At the unitarity limit for $\alpha=1$ we obtain
 $\hat{\mu}_{\downarrow}=-0.535$, which should be compared with
 $\hat{\mu}_{\downarrow}=-0.57$ \cite{Schmidt:2011zu} from BMW-type FRG,
 $\hat{\mu}_{\downarrow}=-0.618$ \cite{Prokofiev2008a} and $-0.615$
 \cite{Prokofiev2008b,Vlietinck1302} from the diagMC, 
 $\hat{\mu}_{\downarrow}=-0.6066$ from a variational method \cite{Combescot0702}, 
 and $\hat{\mu}_{\downarrow}=-0.6156$ from \cite{Combescot0804}. 
 Furthermore, at $1/(k_Fa)=1.0$ with $\alpha=1$,
 we obtain $\hat{\mu}_{\downarrow}=-2.33$, which lies slightly above the
 value $-2.62$ from \cite{Prokofiev2008a,Prokofiev2008b,Vlietinck1302}.
 In the case of a heavy impurity ($\alpha=6.6$) at unitarity, we obtain
 $\hat{\mu}_{\downarrow}=-0.385$, not far away from the value $-0.44$
 from the $T$-matrix approach \cite{Baarsma1110}.  In the case of a
 light impurity ($\alpha=0.5$) at unitarity, we obtain
 $\hat{\mu}_{\downarrow}=-0.772$, to be compared with $\sim -0.9$ from
 the diagMC \cite{Kroiss1411} and $\sim -0.87$ from a variational method
 \cite{Combescot0702}.

 \begin{figure}[bt]
  \begin{center}
   \includegraphics[width=0.95\columnwidth]{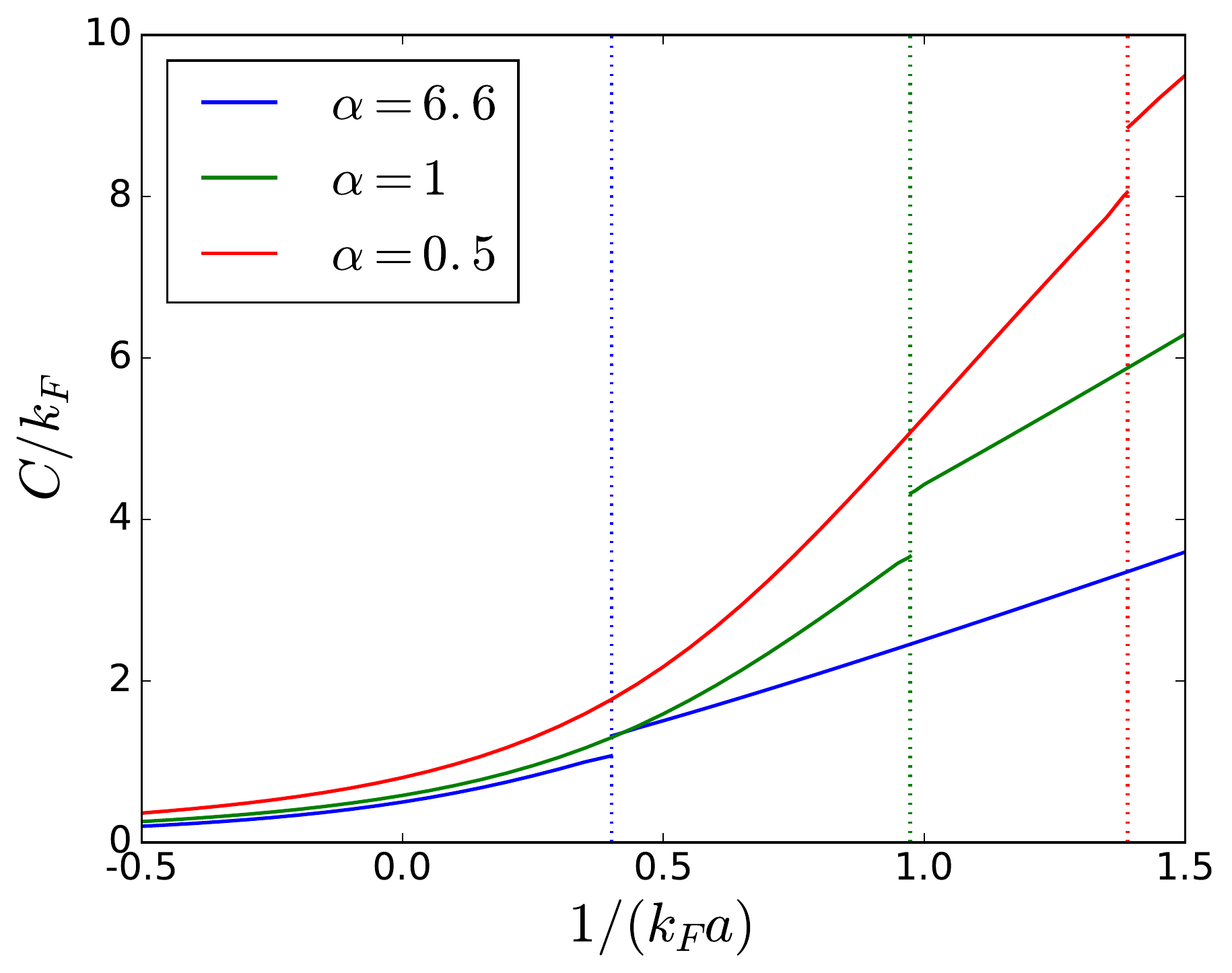}
  \end{center}
  \vspace{-1.5\baselineskip}
  \caption{\label{fig:Tan_contact} Tan's contact for $\alpha=0.5$ (upper curve),
    $\alpha=1$ (middle curve), and $alpha=6.6$ (lower curve).
    Vertical dotted lines are the same as in
  Fig.~\ref{fig:ground_state_energy}.}
 \end{figure}
 Based on the above evaluation, we can determine the so-called Tan's
 contact, which is directly related to the slope of the ground state
 energy:
 \ba
	  \begin{split}
		   C &= 8 \pi M_r a^2 \frac{\dd E}{\dd a}
		   \\
		   & = - 8 \pi M_r  k_F \frac{\dd \hat\mu_{\down}^c}{\dd[1/(k_Fa)]}
	  \end{split}
	  \label{eq:tanC}
 \ea
 where $M_r=(M_\up^{-1}+M_\down^{-1})^{-1}$ is the reduced mass.  We
 refer to Appendix \ref{ap:Tancontact} for more details on
 \eqref{eq:tanC}.  In Fig.~\ref{fig:Tan_contact}, Tan's contact in a
 mobile impurity system is plotted. It clearly shows the first-order
 nature of the transition: since the slope of the ground state energy is
 discontinuous at the polaron-molecule transition point, Tan's contact
 also shows a discontinuity \cite{Punk0908}.  A measurement of Tan's
 contact may be useful to determine the polaron-molecule transition
 experimentally.

 \subsection{Quasiparticle properties}

 Next we discuss spectral properties of the impurity.
 \begin{figure}[bt]
	  \begin{center}
	  	\includegraphics[width=0.95\columnwidth]{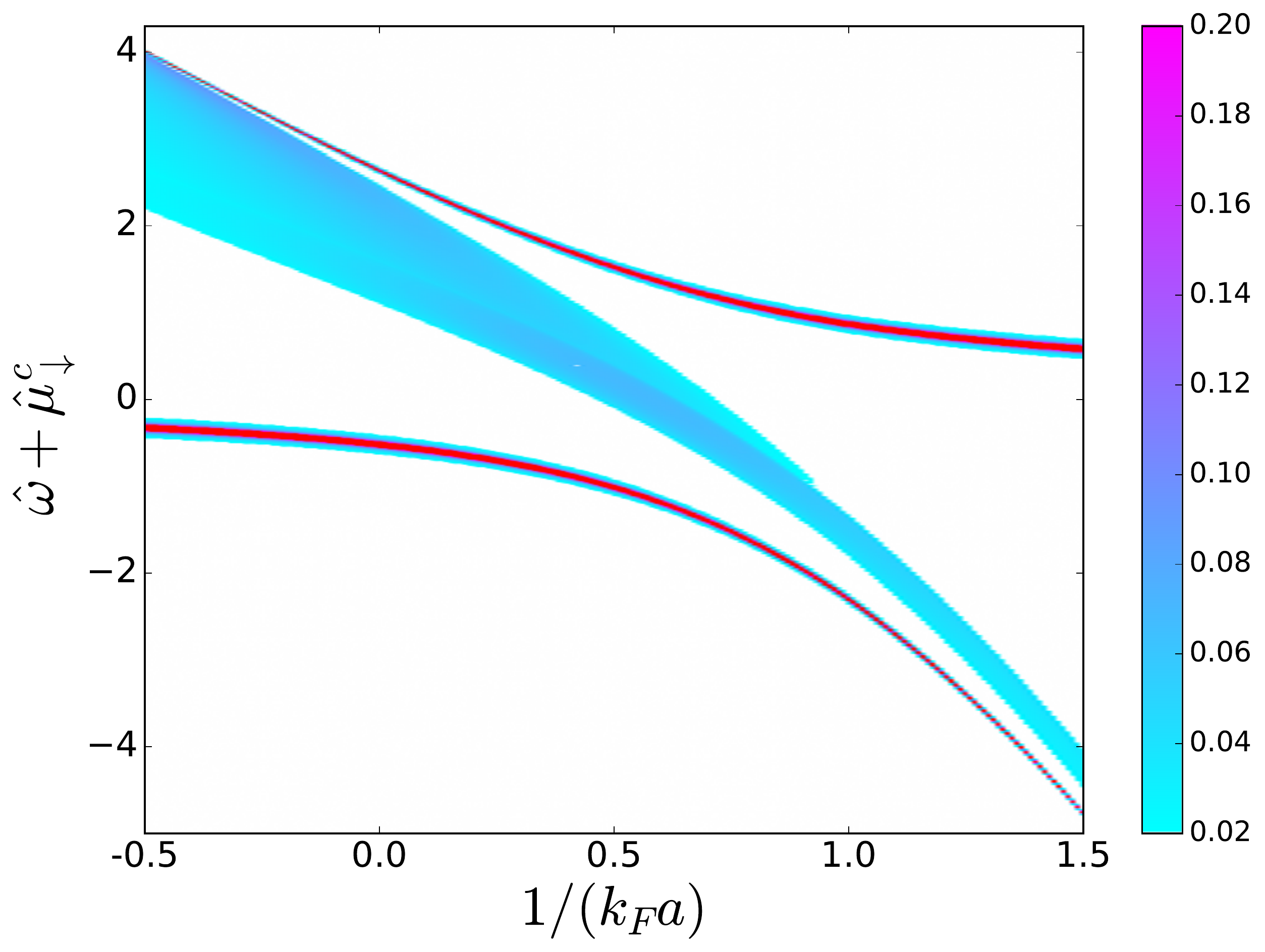}
	  \end{center}
	  \vspace{-\baselineskip}
	  \caption{\label{fig:spectrum_d_alpha=1}
	  Spectral density of the polaron at zero momentum 
	  $\A_{\down}(\omega,\mathbf{p}=\mathbf{0})$ 
	  for $\alpha=1$.    
	  $\A_{\down} < 0.02$ ($\A_{\down}>0.2$) 
	  in the white (red) region, respectively.}  
 \end{figure}
 In Fig.~\ref{fig:spectrum_d_alpha=1} we plot the spectral density of
 the polaron in the equal-mass system.  There are two quasiparticle
 peaks \cite{CuiZhai1001}: the high-energy branch, called the \emph{repulsive polaron},
 and the lower-energy branch called the \emph{attractive polaron}.
 Between these two peaks, there appears the molecule-hole
 continuum. These spectral features are consistent with previous studies
 \cite{Massignan1102,Schmidt:2011zu,Goulko2016a}.  We have also examined
 the spectral density in a mass-imbalanced system and found that it
 shares the same structure as in Fig.~\ref{fig:spectrum_d_alpha=1}.

 \begin{figure}[htbp]
  \begin{minipage}{0.49\columnwidth}
   \begin{center}
    \includegraphics[width=0.99\columnwidth]{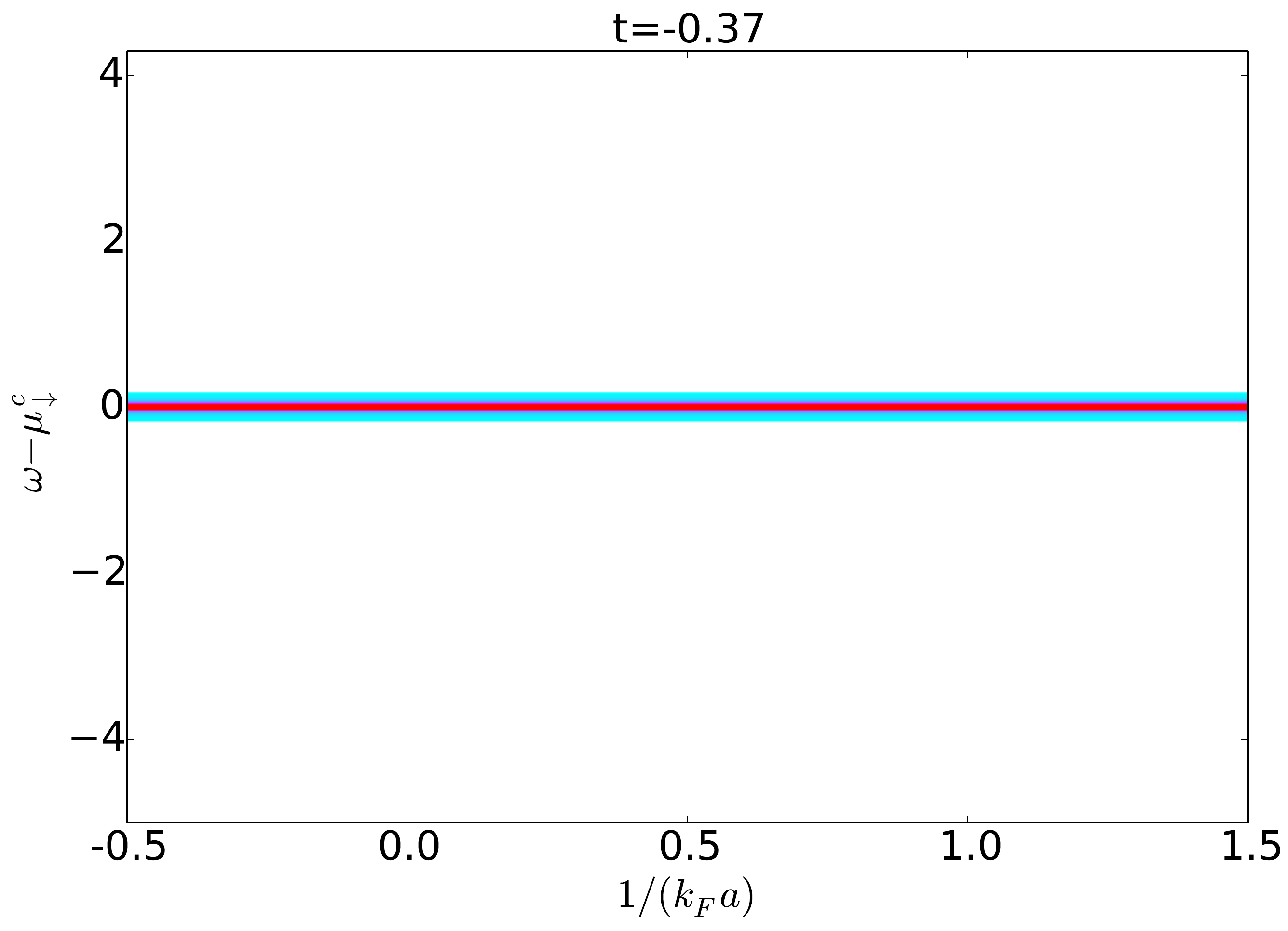}
   \end{center}
  \end{minipage}
  \begin{minipage}{0.49\columnwidth}
    \begin{center}
     \includegraphics[width=0.99\columnwidth]{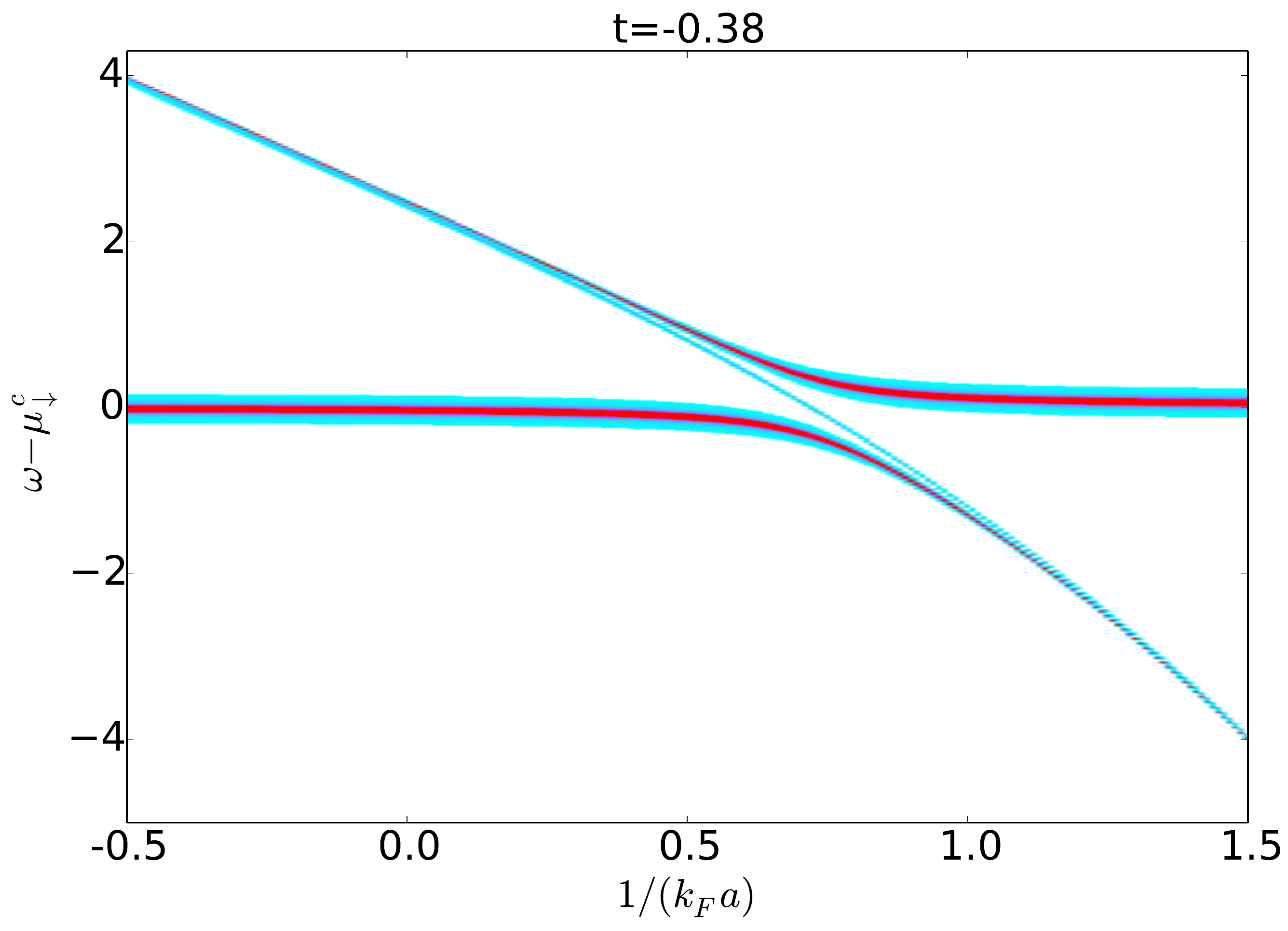}
    \end{center}
  \end{minipage}
  \begin{minipage}{0.49\columnwidth}
    \begin{center}
     \includegraphics[width=0.99\columnwidth]{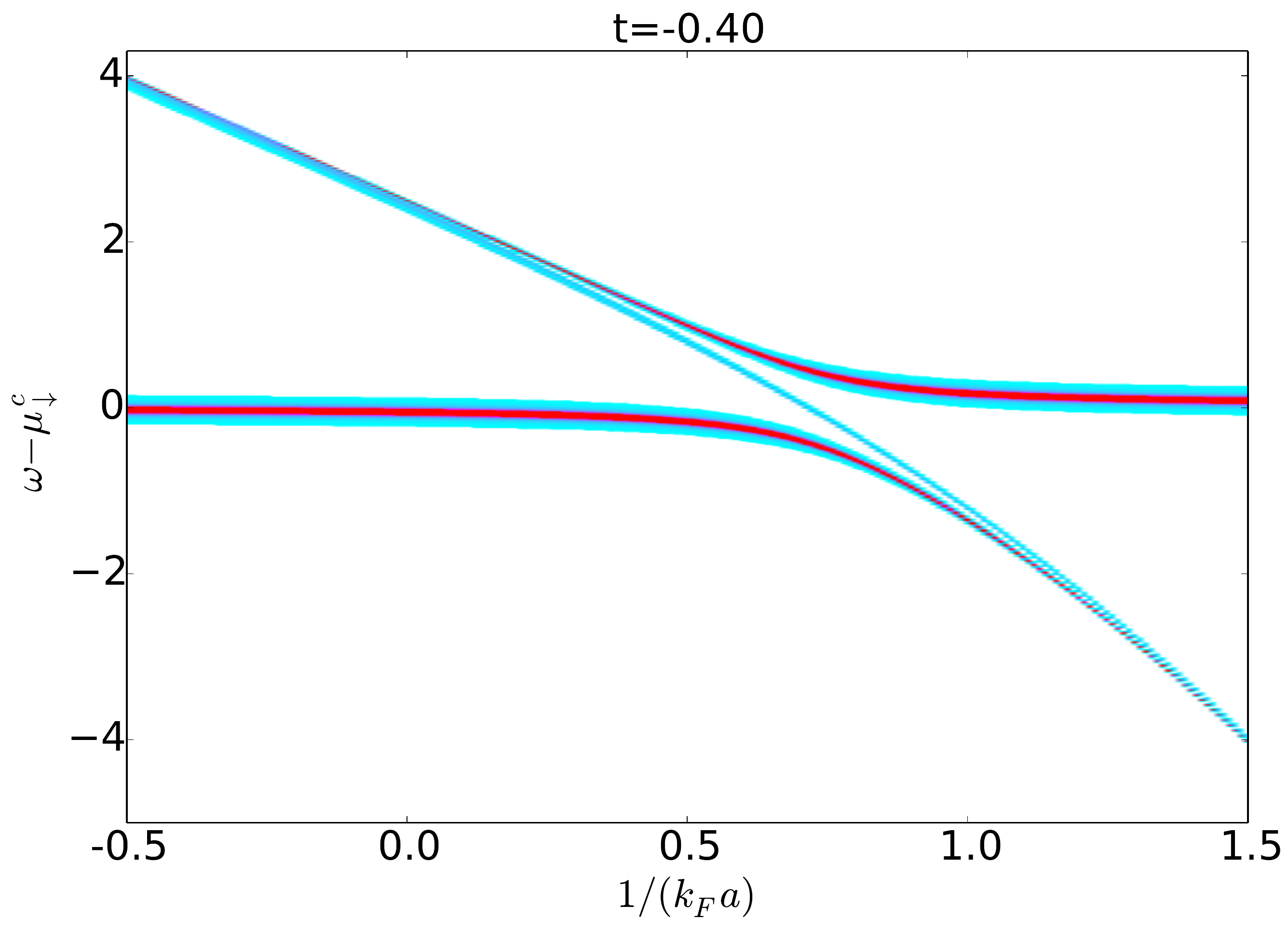}
    \end{center}
  \end{minipage}
  \begin{minipage}{0.49\columnwidth}
    \begin{center}
     \includegraphics[width=0.99\columnwidth]{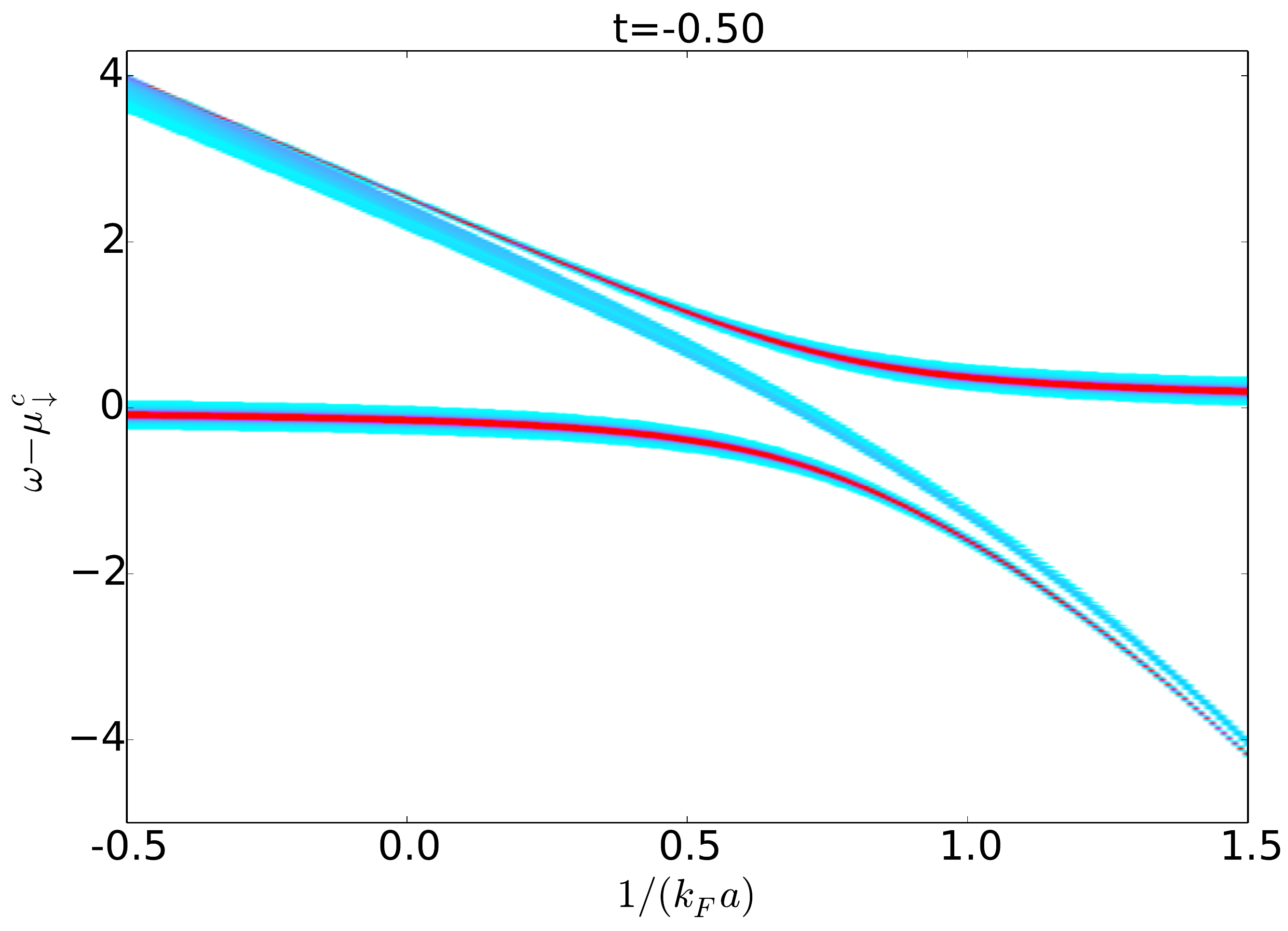}
    \end{center}
  \end{minipage}
  \begin{minipage}{0.49\columnwidth}
    \begin{center}
     \includegraphics[width=0.99\columnwidth]{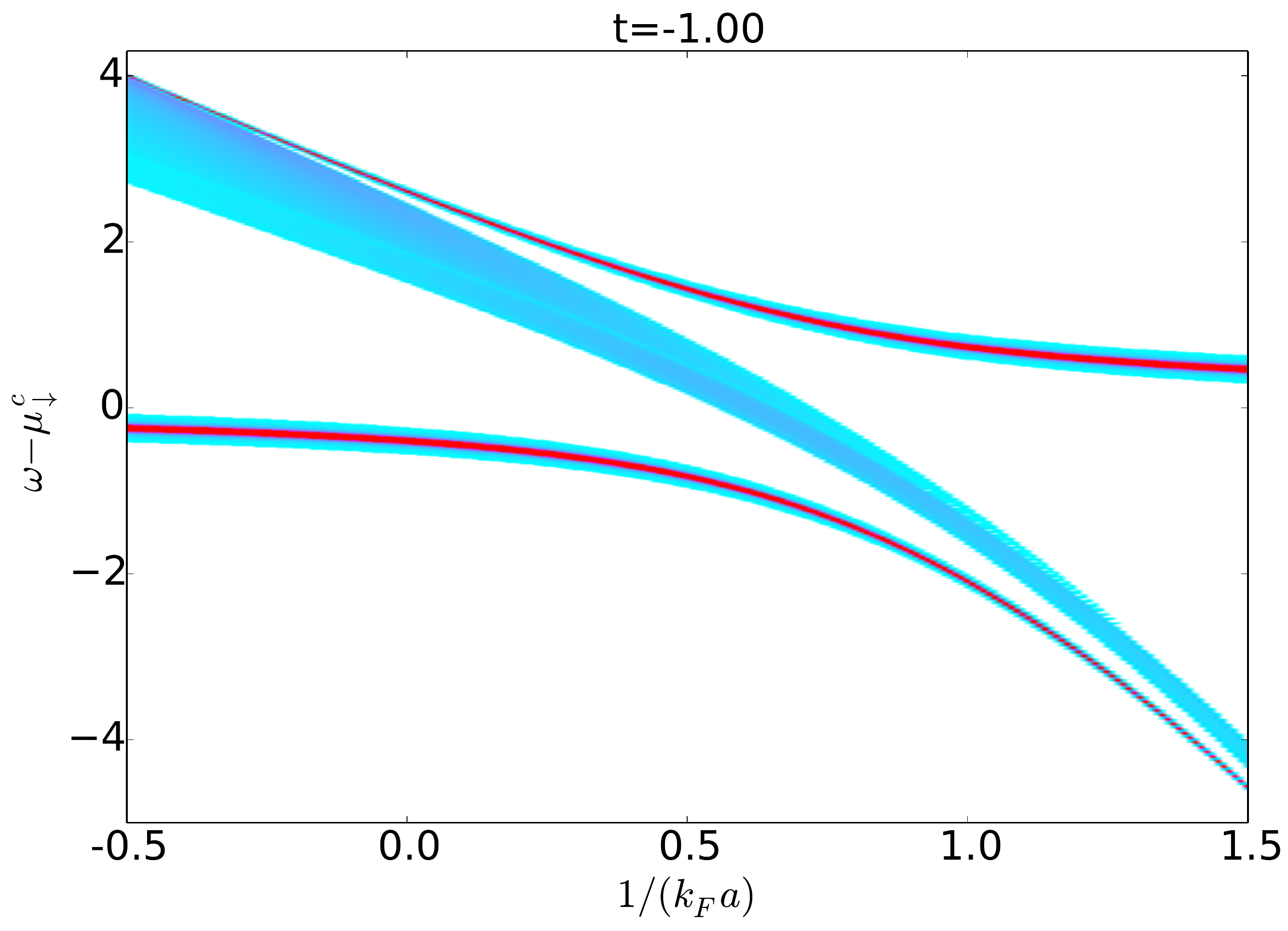}
    \end{center}
  \end{minipage}
  \begin{minipage}{0.49\columnwidth}
    \begin{center}
     \includegraphics[width=0.99\columnwidth]{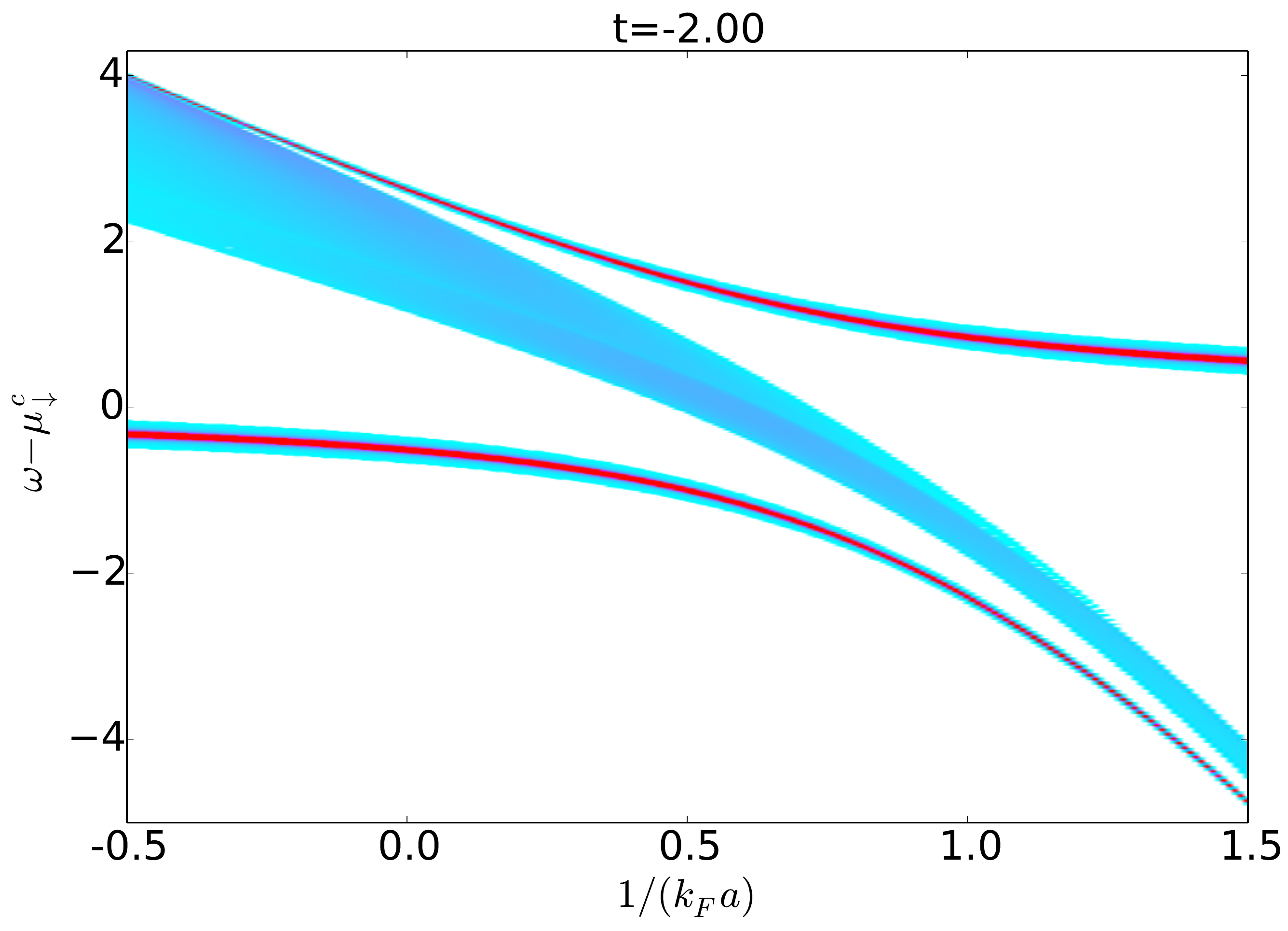}
    \end{center}
  \end{minipage}
  \caption{\label{fig:scale_dependent_spectral} Scale evolution of
  $\A_{\down}(\omega,\mathbf{p}=\mathbf{0})$ for $\alpha=1$. 
  The color map is the same as in Fig.~\ref{fig:spectrum_d_alpha=1}.}
 \end{figure}
 In Fig.~\ref{fig:scale_dependent_spectral}, we show the RG scale
 dependence of the spectral density of the polaron in the equal-mass
 system. For $t\geq -0.37$, the spectral density remains trivial or
 classical. At $t=-0.38$, the horizontal branch suddenly splits into two
 branches and a shallow continuum in between. After $t=-0.38$ the two
 peaks are gradually deformed to the forms in the IR limit, while the
 molecule-hole continuum increases the width especially on the polaron
 side $1/(k_Fa)\lesssim 1$. It is intriguing that both the repulsive
 polaron and the attractive polaron originate partly from the trivial
 unrenormalized spectrum in the UV.

 \begin{figure}[h]
	  \centering
	  \includegraphics[width=0.95\columnwidth]{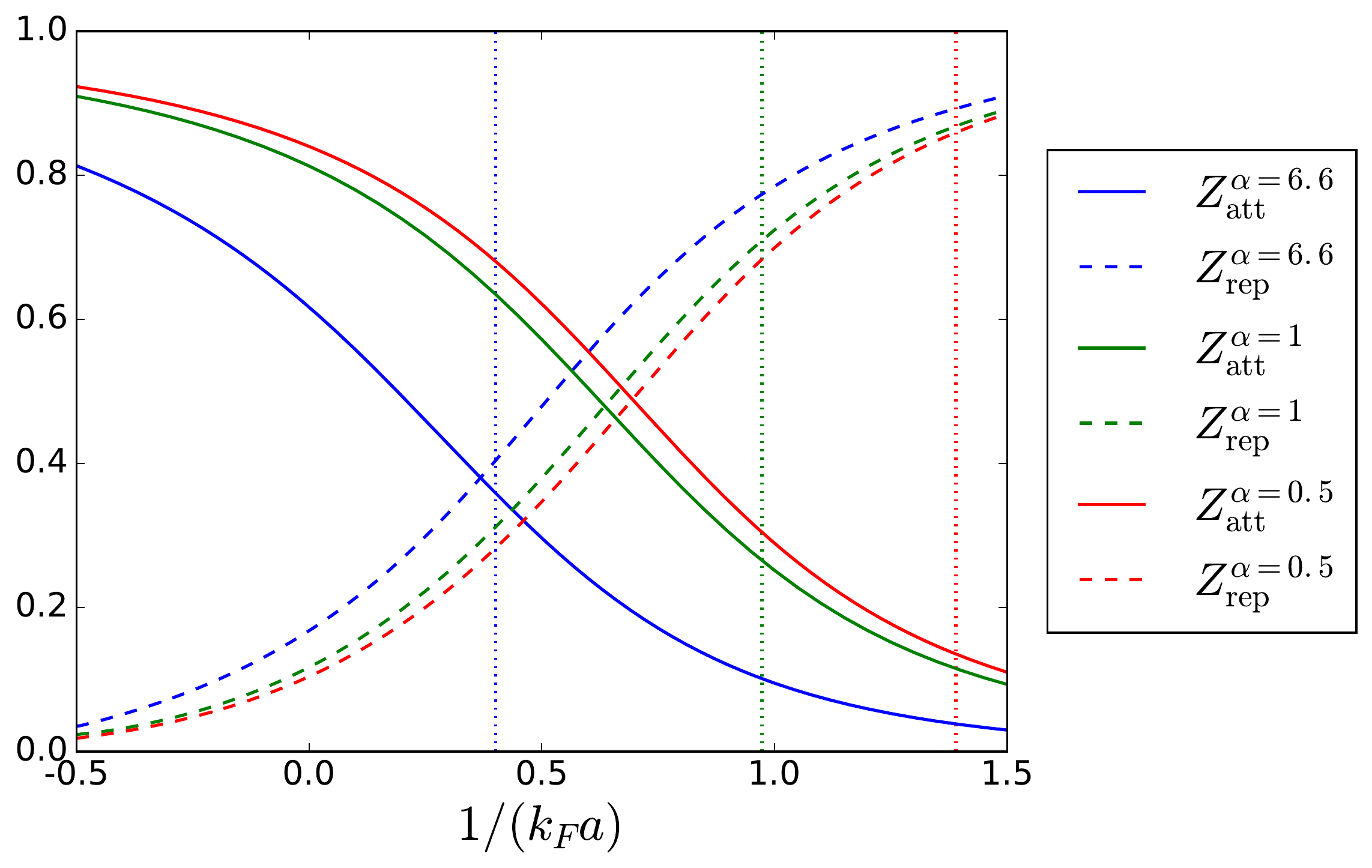}
	  \vspace{-0.5\baselineskip}
	  \caption{\label{fig:weight}
	  Quasiparticle weight of the impurity for the attractive and
	  repulsive branches.
          The upper (lower dotted), middle (dotted), and lower (upper dotted)
          curves are respectively the results for $\alpha=0.5$, $\alpha=1$,
          and $\alpha=6.6$.
          The dotted vertical lines represent the  
	  polaron-molecule transition for each $\alpha$.}
 \end{figure}
 The quasiparticle weights are calculated from the zeros of the real
 part of the two-point function in the IR limit. First we numerically 
 determine the peak positions of the attractive and repulsive polaron,
 \ba
	{\rm Re}~\P^{\rm R}_{\down}(\omega, \mathbf{p}=\mathbf{0}) 
	\Big|_{\omega = \omega^{\mathstrut}_{\rm att/rep}} 
	\!\! = 0\,. 
 \ea
 The energy of the repulsive polaron determined this way was observed to 
 converge to the asymptotic formula \cite{Bishop1973391} for small positive $k_Fa$.   
 At the zeros, the quasiparticle weights are determined as
\ba
	Z_{{\rm att}/{\rm rep}} \equiv
	- \left[ \partial_{\omega} {\rm Re}~\P^{\rm R}_{\down}
	(\omega, \mathbf{p}=\mathbf{0}) 
	\right]^{-1}\Big|_{\omega = \omega^{\mathstrut}_{\rm att/rep}} .
\ea
 The above definition implicitly relies on the fact that the imaginary
 part of the ${\bf p}={\bf 0}$ two-point function at the zeros are
 vanishingly small in the present scheme at $\hat\epsilon\ll 1$. (If the
 imaginary part is not small, the definition above is no longer valid
 and one needs to search for a quasiparticle pole on the complex
 $\omega$ plane.)  Considering that the repulsive polaron is a
 high-energy metastable branch and naturally has a finite life time
 \cite{CuiZhai1001,Massignan1102,Schmidt:2011zu}, the absence of the decay width may
 be an artifact of the present scheme.
 
 In Fig.~\ref{fig:weight}, we plot the quasiparticle weight of the
 attractive and repulsive polarons.  On the molecule side, the repulsive
 polaron is dominant for every mass ratio. When the scattering length is
 shifted to the polaronic side, the weight of the repulsive polaron
 decreases and exchanges dominance with the attractive polaron. Note
 that the crossing of $Z_{\rm rep}$ and $Z_{\rm att}$ does not exactly
 occur at the polaron-molecule transition. These features of the weight
 $Z$ are consistent with previous studies
 \cite{Punk0908,Massignan1102,Schmidt:2011zu,Massignan1112,Vlietinck1302}.
 Quantitatively, for equal masses ($\alpha=1$) at unitarity we obtain
 $Z_{\rm att}=0.813$, which is close to the values $0.78\sim 0.80$ from
 the variational method \cite{Punk0908}, the $T$-matrix method
 \cite{Massignan1112} and the BMW-type FRG \cite{Schmidt:2011zu}, as
 well as $0.7586(27)$ from the diagMC \cite{Vlietinck1302}. Note however
 that all of these results are well above the value $0.39(9)$ obtained
 in an experiment on $^6$Li atoms \cite{Schirotzek0902}.  In the case of
 a heavy impurity ($\alpha=6.6$) at unitarity we obtain $Z_{\rm
 att}=0.617$, which is close to the value $\sim 0.64$ from the
 $T$-matrix method \cite{Baarsma1110} and the diagMC \cite{Kroiss1411}.
 
 \begin{figure}[bt]
	  \centering 
	  \includegraphics[width=0.95\columnwidth]{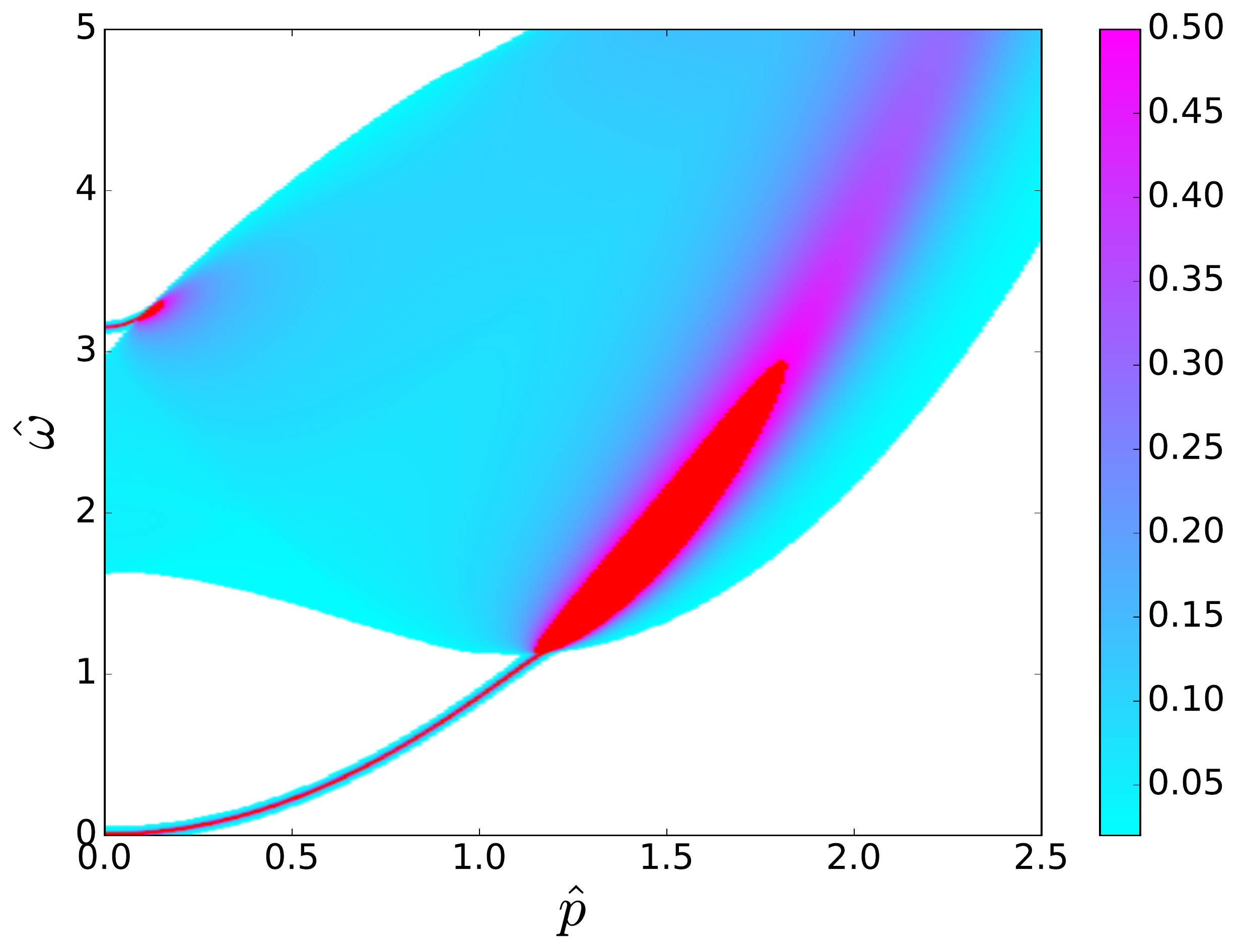}
	  \vspace{-0.5\baselineskip}
	  \caption{\label{fig:spectrum_momentum_d_alpha=1}
	  $\A_{\down}(\omega, \mathbf{p})$ for $\alpha = 1$ in the unitarity limit. 
	  $\A_{\down} < 0.02$ ($\A_{\down}>0.5$) in the white (red) region, respectively.}
 \end{figure}
 Next we discuss the nature of the quasiparticles at ${\bf p}\ne{\bf
 0}$. The polaron spectral density at finite momenta in the unitarity
 limit is shown in Fig.~\ref{fig:spectrum_momentum_d_alpha=1}. At zero
 momentum, neither the attractive nor the repulsive polaron branch has a
 width. By contrast, at finite momentum, the peaks merge with the
 molecule-hole continuum and acquire a finite width. As the momentum
 increases, the branches lose sharp peaks and can no longer be clearly
 distinguished from the smooth continuum.

 From the finite-momentum real-time correlation functions, we can
 extract the effective mass of the attractive and repulsive polarons:
 $m_{\down \rm att}$ and $m_{\down \rm rep}$. First we determine the
 dispersion relation of the quasiparticles $E_{\rm att/rep}({\bf p})$ by
 solving the equation
 \ba
 	{\rm Re}~\P^{\rm R}_{\down}(\omega, \mathbf{p}) 
 	\Big|_{\omega = E_{\rm att/ rep}({\bf p})} = 0\,.
 \ea
 Then we fit the dispersion relation with the following fit function,
 \ba
 	E_{\rm att/rep}({\bf p}) - E_{\rm att/rep}({\bf 0}) 
 	= \frac{{\bf p}^2}{2 m_{\down \rm att/rep}} 
 	+ \calO(p^4)\;.
 \ea
 
 \begin{figure}[tbh]
	  \begin{center}
	  	\includegraphics[width=0.95\columnwidth]{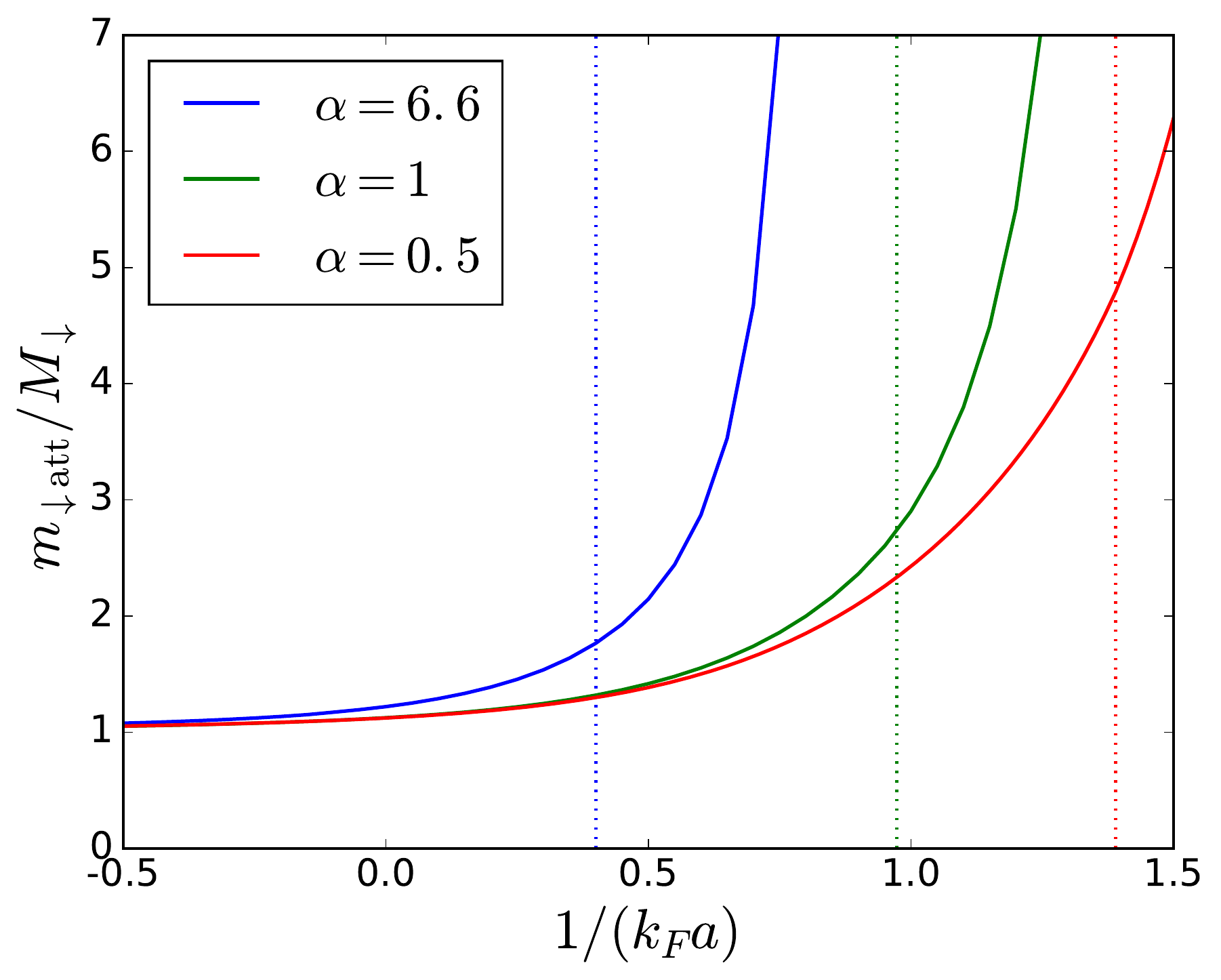}
	  \end{center}
	  \vspace{-1.5\baselineskip}
	  \caption{\label{fig:dispersion_attractive}
	  Effective mass of the attractive polaron in units of the bare 
	  mass $M_\down=\alpha/2$ for $\alpha=6.6$ (upper curve),
          $alpha=1$ (middle curve), and $\alpha-0.5$ (lower curve).
          The dotted
	  vertical line represents the polaron-molecule transition 
	  for each $\alpha$.}
 \end{figure}
 In Fig.~\ref{fig:dispersion_attractive}, the effective mass of the
 attractive polaron is shown. At $(k_Fa)^{-1}=-0.5$, $m_{\down {\rm
 att}}$ for three $\alpha$'s are already close to the asymptotic value
 $M_\down=\alpha/2$ in the free limit $k_F a\to 0^-$. For equal masses
 ($\alpha=1$) at unitarity we find $m_{\down {\rm att}}/M_\down=1.125$,
 which lies above 1.04(3) \cite{Lobo2006} and 1.09(2) \cite{Pilati0710}
 from fixed-node diffusion Monte Carlo simulations, but slightly below
 $1.17$ \cite{Combescot0702} and $1.197$ \cite{Combescot0804} from
 variational methods and $\sim 1.18$ from a $T$-matrix approach
 \cite{Massignan1112}.  For equal masses at $1/(k_Fa)=1.0$, we find
 $m_{\down {\rm att}}/M_\down=2.90$, which is between the value $\sim
 2.5$ from a variational method \cite{Combescot0702} and the value $\sim
 3.5$ from the \mbox{diagMC} \cite{Prokofiev2008a,Prokofiev2008b}.  
 In the case of a heavy impurity 
 ($\alpha=6.6$) at unitarity, we obtain $m_{\down {\rm
 att}}/M_\down=1.22$, whereas the $T$-matrix approach yields $1.16$
 \cite{Baarsma1110}.  In the case of a light impurity at $1/(k_Fa)=1.0$,
 we find $m_{\down {\rm att}}/M_\down=2.43$, while \cite{Combescot0907}
 reports $\sim 3.3$.
 
 At fixed $1/(k_Fa)$, we found $m_{\down {\rm att}}/M_\down$ to be an
 increasing function of $\alpha$, which agrees with \cite{Combescot0907}
 but disagrees with \cite{Combescot0702}.  Overall, the effective mass
 monotonically increases across the polaron-molecule transition and
 eventually diverges at some point.  Deep in the molecular region, we
 found $m_{\down{\rm att}}<0$ (not shown in
 Fig.~\ref{fig:dispersion_attractive}), in agreement with
 \cite{Trefzger1112}; this is consistent with the fact that the polaron
 is not a stable quasiparticle in this region.

 We also calculated the effective mass of the repulsive polaron, finding
 that although $m_{\down \rm rep}$ expectedly converges to the bare mass
 $\alpha/2$ at $1/(k_Fa)\gg 1$, it \emph{decreases} to 0 near the
 unitarity limit.  The latter behavior contradicts other studies
 \cite{Massignan1102,Massignan1112} showing an \emph{enhancement} of
 $m_{\down \rm rep}$ towards unitarity.  This could be indicating 
 a shortcoming of the present scheme in a strongly-interacting regime.
 For a small negative $1/(k_Fa)$, we found that the repulsive polaron
 branch merges with the molecule-hole continuum even at small momenta,
 which makes $m_{\down \rm rep}$ ill-defined in this limit.

\section{\label{sc:disc}Summary}

We have examined a mobile impurity immersed in a Fermi sea realized in
cold atoms.  We have calculated thermodynamic quantities by means of the
derivative expansion method in FRG and obtained results that are in
reasonable agreement with the diagMC simulations and other analytical
methods.

The important aspect of the mobile impurity problem is that the spectral
properties are rich since in addition to the attractive polaron, the
molecule-hole continuum and the repulsive polaron appear in higher
frequency regimes.  To extract spectral properties within FRG, we have
adopted a real-time method originally developed in nuclear physics and
showed that it is successful in generating the spectral weights
at high frequencies that are difficult to obtain within the conventional
derivative expansion method.
The key point is solving the flow equations (Eq.~\eqref{eq:rfd}) in an iterative manner.
Our analyses show that one iteration allows one to reproduce 
qualitative features in the Fermi polaron system.
A quantitative agreement with the diagMC method may be
obtained by additional iterations,
whose implementation is an interesting future work.
Applying our method to the other
experimentally-achievable systems such as the
Bose Polaron~\cite{Hu2016,Jorgensen2016},
the BCS-BEC crossover in two-component Fermi gases~\cite{Zwerger2011},
and the Bose-Fermi mixture systems~\cite{Modugno2002,Ospelkaus2006}
may also be interesting.

The polaron problem may be one of the best systems to analyze the
accuracy and potential of a many-body technique thanks to its
experimental accessibility and applicability of the diagMC method,
in marked contrast to a number of models discussed in nuclear and
particle physics.  Our consideration may pave the way to an efficient
algorithm in real-time FRG.

\begin{acknowledgments}
  T.~K. was supported by the RIKEN iTHES project. 
  T.~K. thanks P.~Massignan for a useful correspondence. 
  S.~U. thanks T. Enss for useful discussions.
  S.~U. was supported by Grant for Basic Science
  Research Projects from The Sumitomo Foundation.
\end{acknowledgments}

\appendix
\section{\label{ap:2pfloweq}Flow equations for 2-point functions}
By means of the integration formula \eqref{eq:3dintformula}, the flow
equations \eqref{eq:fl0} are recast into the form
\begin{widetext}
\ba
	\der_t \hat{\P}^{\rm R}_{\down}(\hat{\omega},\hat{p}) 
	& =  
	\frac{\hat{h}^2 \ee^{2t}}{4\pi^2 A_{\phi,k}}
	\Bigg\{
        \ee^{t}\Gamma\Big( 1-2\ee^{2t}-\hat{p}^2, ~
        2\hat{p} \ee^{t},~
        1 - \hat{p}^2-\frac{\alpha}{1+\alpha}\ee^{2t}
        +\frac{\hat{m}^{2}_{\phi,k}}{
		A_{\phi,k}} - \hat{\omega} - i\hat{\epsilon}\Big)
	\nonumber
	\\
	&  \qquad + \theta(1 - \ee^{2t})\ 
	    \sqrt{1 - \ee^{2t}}\ 
          \Gamma\Big(\frac{1-2\ee^{2t}+\hat{p}^2}{1+\alpha}, ~
          \frac{2 \hat{p}\sqrt{1 - \ee^{2t}}}{1+\alpha}, ~ 
          \frac{1+\hat{p}^2+\alpha \ee^{2t}}{1+\alpha}
          + \frac{\hat{m}^{2}_{\phi,k}}{
            A_{\phi,k}}-\hat{\omega}-i\hat{\epsilon}\Big)
	\Bigg\}\,,
	\\
	\der_t \hat{\P}^{\rm R}_{\phi}(\hat{\omega},\hat{p}) 
        & =  \frac{\hat{h}^2 \ee^{2t}}{4\pi^2 A_{\down,k}}\Bigg\{
          \ee^{t}\Gamma\Big(
	          \hat{p}^2- 1, ~ -2\hat{p} \ee^{t},~
	          \hat{p}^2- 1+\frac{1+\alpha}{\alpha} \ee^{2t}
	          +\frac{\hat{m}^{2}_{\down,k}}{
	            A_{\down,k}} - \hat{\omega} - i\hat{\epsilon}
		\Big)
		\nonumber
		\\
        & \qquad 
         + \sqrt{1 + \ee^{2t}}\ \Gamma\Big(
          \frac{1+\hat{p}^2}{\alpha},~
          \frac{2\hat{p}\sqrt{1+\ee^{2t}}}{\alpha},~
          \frac{1+\hat{p}^2+(1+\alpha) \ee^{2t}}{\alpha}
          +\frac{\hat{m}^{2}_{\down,k}}{A_{\down,k}} 
          - \hat{\omega} - i\hat{\epsilon}
	     \Big)
        \Bigg\}\,,
\ea
\end{widetext}
where the analytic continuation \eqref{eq:ancont} has been performed. 
The above flow equations are solved under the initial conditions
\ba
	\hat{\P}^{\rm R}_{\down,\Lambda}(\hat{\omega},\hat{p}) 
	& = 
	A_{\down,\Lambda}(-\hat{\omega}-i\hat{\epsilon}+\hat{p}^2/\alpha) 
	+ \hat{m}_{\down,\Lambda}^2 \,,
	\\
	\hat{\P}^{\rm R}_{\phi,\Lambda}(\hat{\omega},\hat{p}) 
	& = A_{\phi,\Lambda}[-\hat{\omega}-i\hat{\epsilon}+\hat{p}^2/(1+\alpha)] 
	+ \hat{m}_{\phi,\Lambda}^2 \,.
	\!\!
\ea

\section{\label{ap:vaclim}Vacuum limit}

In this appendix we derive initial conditions at $k=\Lambda$ for flow 
equations in the derivative expansion. The guiding principle here is to adjust 
the initial parameters so as to reproduce the vacuum two-body scattering correctly 
\cite{Diehl:2007ri,Diehl:2009ma,Boettcher:2012cm}. The initial conditions 
for $\alpha=1$ have already been discussed in the Appendix of \cite{Schmidt:2011zu} 
and we aim here to extend their results to general $\alpha$. 

\begin{figure}[b]
	\centering
	\includegraphics[width=.4\textwidth]{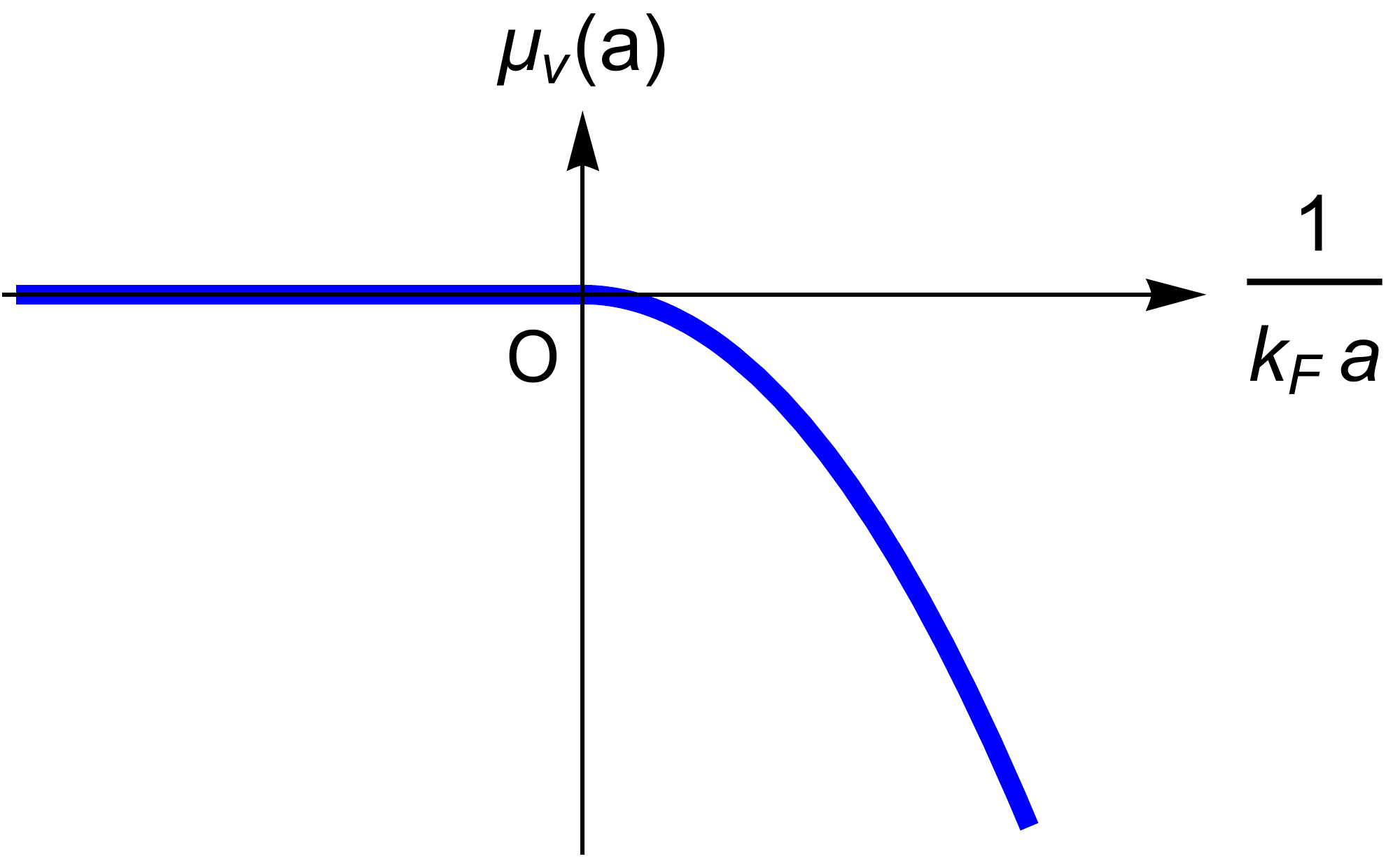}
	\caption{\label{fg:mu_v} 
	The $a$-dependence of $\muv(a)$.}
\end{figure}

Let us consider a two-body system of $\up$ and $\down$ atoms in vacuum. In the present 
formalism of FRG, this ``vacuum limit'' is not realized by the naive choice 
$\mu_\up=\mu_\down=0$, because 
atoms are gapless only for $a<0$ (BCS side).  On the BEC side where $a>0$, 
molecules are gapless while atoms are \emph{gapped} due to a bound state formation. 
This means that the vacuum limit for $a>0$ must be reached via a flow equation with 
$\mu=\muv(a)<0$ for atoms 
\cite{Diehl:2005ae,Diehl:2007th,Diehl:2007ri,Diehl:2007xz,Diehl:2009ma,Boettcher:2012cm}, 
where $\muv(a)$ should be tuned with $a$ (Fig.~\ref{fg:mu_v}).   

Let us therefore start with $\mu_\up=\mu_\down=\muv(a)$. 
Since there is no flow of $A_\down$ and $m_\down^2$ in this case, 
one can let $A_{\down}=1$ and $m^2_{\down}=
-\muv(a)$ for all $k$. The effective action in the derivative expansion at the 
microscopic scale $k=\Lambda$ is then given by
\ba
	\label{eq:ginit}
	\Gamma_{k=\Lambda} & = \int_{P} 
	\Bigg\{\displaystyle 
		\sum_{\sigma=\up,\down}\psi_\sigma^*\Big[-i p_0+
		\frac{\mathbf{p}^2}{2M_\sigma}
           - \muv(a)\Big]\psi_\sigma
	\notag
	\\ 
	&  ~~ + \phi^* \Big( A_{\phi,\Lambda}\Big[ -i p_0+
			\frac{\mathbf{p}^2}{2M_\phi} - 2\muv(a) \Big] 
			+ \mm^2_{\phi,\Lambda}(a) \Big)\phi
	\Bigg\}
	\notag
	\\
	& ~~ 
	+ h \int_{\mathbf{x},\tau}(\psi_\up^*\psi_\down^*\phi+\text{H.c.})\,,
\ea
where we have introduced a ``mass term'' $\mm^2_{\phi,\Lambda}$ for $\phi$. 
It follows that 
\ba
	m^2_{\phi,\Lambda} = \mm^2_{\phi}(a) 
	- 2 A_{\phi,\Lambda} \muv(a)\,.
	\label{eq:mini5}
\ea
We need to tune both $\mm_\phi(a)$ and $\muv(a)$ to recover 
correct IR properties in vacuum. To proceed, let us recall the fact that 
it is possible to compute the vacuum molecule propagator \emph{exactly} 
starting from the microscopic action \eqref{eq:ginit} using the method of 
\cite{Gurarie0611,Levinsen1101}, with the result 
(continued to Minkowski spacetime)
\begin{multline}
	\P_{\phi}(\omega,\mathbf{p};\muv) = A_{\phi,\Lambda}
	\Big(
		-\omega+\frac{\mathbf{p}^2}{2M_\phi}-2\muv
	\Big)+\mm_\phi^2(a)
	\\
	- \frac{M_r}{\pi^2}h^2\Lambda
	+ \frac{M_r^{3/2}}{\pi}h^2\sqrt{
		- \frac{\omega}{2}+\frac{\mathbf{p}^2}{4M_\phi}-\muv
	}\,,
	\label{eq:exactPphi}
\end{multline}
where $\displaystyle M_r\equiv(M_{\up}+M_{\down})^{-1}
=\frac{\alpha}{2(1+\alpha)}$ is the reduced mass, and
a sharp cutoff $\Lambda$ on the spatial momentum was used 
to regularize a UV divergence. 
The parameters in \eqref{eq:exactPphi} can be related to the 
$s$-wave scattering length $a$ and the effective range $r_e$ along the lines of 
\cite{Diehl:2009ma,Boettcher:2012cm} as follows. 
Given the $T$-matrix $T(q)$ for on-shell 
two-body scattering 
$(\mathbf{q},-\mathbf{q})\to(\mathbf{q}',-\mathbf{q}')$ 
in the center-of-mass frame with $|\mathbf{q}|=
|\mathbf{q}'|=q$, there is a relation 
\ba
	T(q) & = - \frac{h^2}
	{\P_{\phi}(-2\muv+\frac{q^2}{2M_r},\mathbf{0};\muv)}\,.
\ea
On the other hand, the general scattering theory defines $a$ and 
$r_e$ through 
\ba
	T(q) & = - \frac{2\pi}{M_r}f(\mathbf{q})
	= -\frac{2\pi}{M_r} \frac{1}{-\frac{1}{a}+\frac{1}{2}r_e q^2-iq}\,.
\ea
Therefore
\ba
	& \frac{M_r}{2\pi}h^2 \Big(-\frac{1}{a}+\frac{1}{2}r_e q^2-iq\Big)
	\notag
	\\
	\overset{!}{=} \, & \P_{\phi}\Big(-2\muv+\frac{q^2}{2M_r},\mathbf{0};\muv\Big)
	\\
	= \, & - \frac{A_{\phi,\Lambda}}{2M_r}q^2 + \mm_\phi^2(a) 
	- \frac{M_r}{\pi^2}h^2\Lambda
	- i \frac{M_r}{2\pi}h^2q\,.
\ea
Comparing each term, we find
\ba
	\mm_\phi^2(a) & = \frac{M_r}{\pi^2}h^2\Lambda
	- \frac{M_r}{2\pi a}h^2 \,,
	\label{eq:mmmf}
	\\
	r_e & = - \frac{2\pi}{M_r^2}\frac{A_{\phi,\Lambda}}{h^2}\,.
	\label{eq:redesu}
\ea
It should be stressed that these relations 
are correct for \emph{both} signs of $a$. Clearly, 
the limit $h\to\infty$ corresponds to a broad FR ($r_e\to 0$).  
We remark that \eqref{eq:mmmf} can also be derived by integrating 
\eqref{eq:mpfll} directly. If we set $A_{\phi,\Lambda}=1$ and  
recall \eqref{eq:mini5}, it follows that we should take 
$m^2_{\phi,\Lambda} = \mm^2_{\phi}(a) -2\muv(a)$ as 
the molecule initial condition \emph{in the vacuum}. 

In order to consider a finite-density system in general, we need to 
tune $\mu_\up$ and $\mu_\down$ away from $\muv(a)$ so that the flow equation 
solved for $k\to 0$ yields the desired number density $n_{\up,\down}$. 
While this is a nontrivial procedure in an interacting system, things become
much simpler in the case of the polaron problem: since 
the $\down$ atoms have a vanishing density, the $\up$ atoms form a 
\emph{noninteracting} Fermi sea with the Fermi momentum $k_F$. 
This means that one may simply set $\mu_\up=k_F^2$ without solving the 
complicated flow equation at all. In summary, the molecule initial condition 
is given by 
\ba
	m^2_{\phi,\Lambda} & = \mm^2_{\phi}(a) - k_F^2 - \mu_\down
	\\
	& = \frac{M_r}{\pi^2}h^2\Lambda - \frac{M_r}{2\pi a}h^2
	- k_F^2 - \mu_\down
	\,.
\ea
This agrees with \eqref{pp4} when converted to the dimensionless units. 

Although the functional form of $\muv(a)$ is not needed in the 
main body of this paper, we wish to include the derivation here for completeness.  
On the BCS side ($a<0$), atoms are gapless physical degrees of freedom, 
so we simply set $\muv(a)=0$. 
On the BEC side ($a>0$), we shall demand instead that 
molecules be gapless. This leads to 
\ba
	0 & \overset{!}{=} \P_{\phi}(0,\mathbf{0};\muv)
	\notag
	\\
	& = -2 A_{\phi,\Lambda}\muv + \mm_\phi^2(a)
	- \frac{M_r}{\pi^2}h^2\Lambda
	+ \frac{M_r^{3/2}}{\pi}
	h^2\sqrt{-\muv}
	\notag
	\\
	& = -2 A_{\phi,\Lambda}\muv 
	- \frac{M_r}{2\pi a}h^2 + \frac{M_r^{3/2}}{\pi}
	h^2\sqrt{-\muv}\,.
\ea
This can be easily solved as
\begin{subequations}
	\ba
		\muv(a) & = - \kappa^2\,, 
		\\
        \kappa & =
	\frac{M_r^{3/2}h^2}{4\pi A_{\phi,\Lambda}}\bigg(
		-1+\sqrt{1+\frac{4\pi A_{\phi,\Lambda}}{M_r^2h^2a}}\,
	\bigg)	\,.
	\ea
\end{subequations}
The $a$-dependence of $\muv(a)$ is illustrated in Fig.~\ref{fg:mu_v}. 
We mention that $\muv(a)$ is equal to half the binding energy of 
the stable dimer that exists at $a>0$. Actually, in terms of $r_e$ 
in \eqref{eq:redesu}, we can write  
\ba
	\kappa & =-\frac{1}{2\sqrt{M_r}\,r_e}
	\bigg(
	-1+\sqrt{1-\frac{2r_e}{a}}\,
	\bigg)\,.
	\label{eq:kp11}%
\ea
which precisely agrees with the well-known formula 
in the effective range model for a narrow FR 
\cite{Gurarie0611,Braaten:2007nq,Levinsen1101}. When $a\gg |r_e|$, we find 
$\kappa\simeq 1/(2\sqrt{M_r}a)$, thus reproducing the universal result 
\ba
	\muv(a) & = - \theta(a)\frac{1}{4M_ra^2}
	\\
	& = - \theta(a) \frac{1+\alpha}{2\alpha}\frac{1}{a^2}
\ea 
for a broad FR.

\section{\label{sec:flowing-parameters}Flowing parameters}
\begin{figure}[b]
	 \centering
	 \includegraphics[width=.4\textwidth]{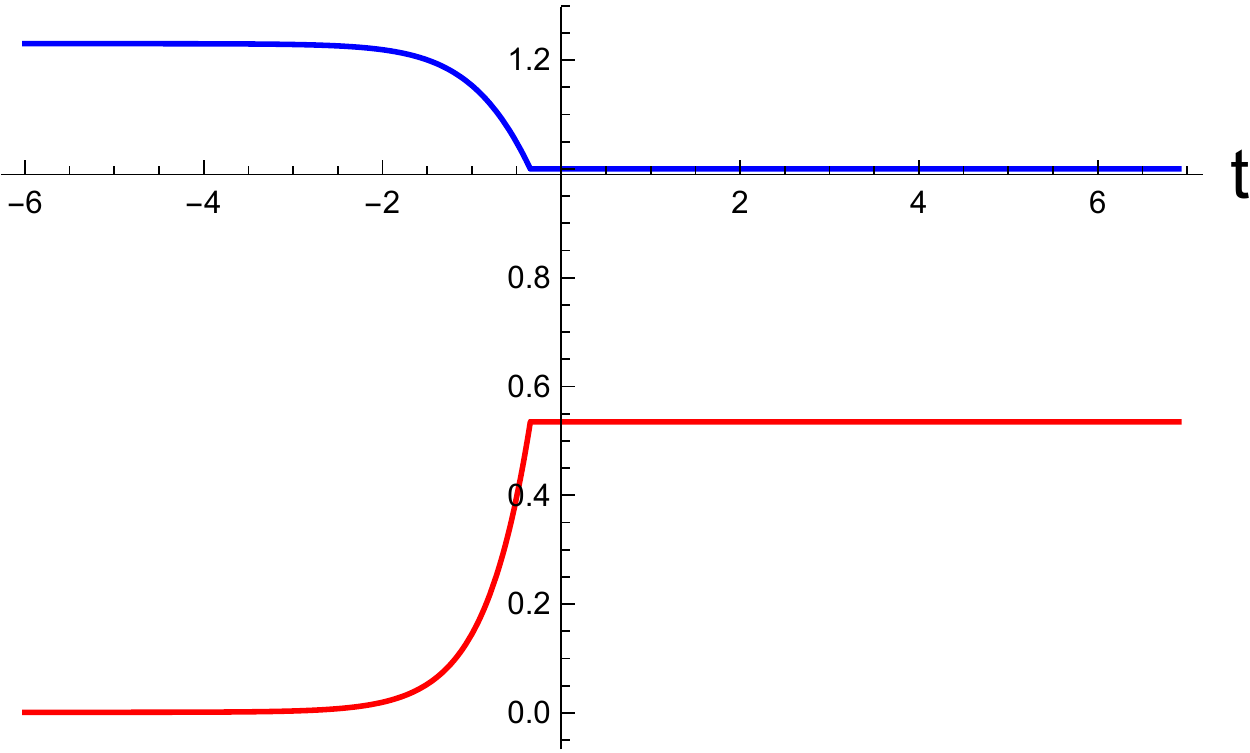}
	 \caption{\label{fg:dd} 
	 Flows \ of $A_{\down}$ (blue curve) and
	 $\hat{m}^{2}_{\down}$
	 (red curve) at unitarity for $\alpha=1$.}
\end{figure}
\begin{figure}[t]
	 \centering
	 \includegraphics[width=.4\textwidth]{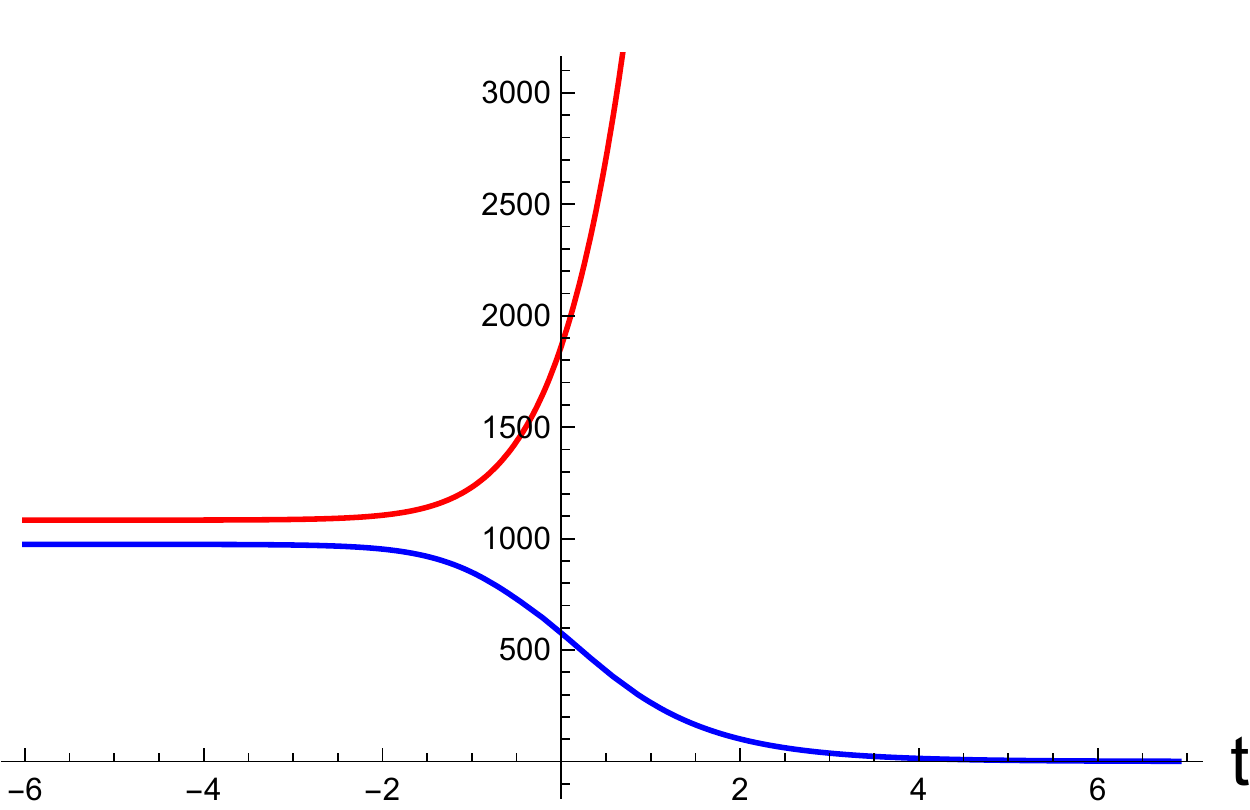}
	 \caption{\label{fg:dp} 
	 Flows of $A_{\phi}$ (blue (upper) curve) and $\hat{m}^{2}_{\phi}$
	 (red (lower) curve) at unitarity for $\alpha=1$.}
\end{figure}
Let us examine the scale dependence of numerical solutions to \eqref{eq:DEfl}.
Figures~\ref{fg:dd} and \ref{fg:dp} show the flows 
in the unitarity limit $[1/(k_Fa)=0]$ 
with $\hat h=300$ and $\alpha=1$. 
As can be seen from the figures,
the flow of each parameter stops at around $t= -4$,
which can therefore be regarded as the $k\to0$ limit.
We also confirmed that the spectral densities have no dependence
on $t_{\text{min}}$ as long as $t_{\text{min}}<-4$.

We observe that $\hat{m}^{2}_{\down}$ flows to zero in the IR limit, 
which indicates that the polaron state is the ground state at unitarity. 
$A_{\down}$ is only slightly renormalized by many-body effects.
By contrast, the renormalization of the molecule is significant; 
$A_{\phi}$ and $\hat{m}^2_{\phi}$ become $\calO(10^3)$ in
 the IR limit.

\section{\label{sec:epsil-depend-spectr}
\boldmath The $\epsilon$-dependence of spectral functions}

In our formulation, analytic continuation 
of  flow equations is performed as $p_0\Rightarrow -i(\omega+i\epsilon)$, 
where $\epsilon$ is an infinitesimal positive constant.  
In numerical implementation, however, $\epsilon$ cannot be made 
arbitrarily small and it is important to understand the 
$\epsilon$-dependence of spectral functions.  As a case study,
we consider $1/(k_Fa)=0.7$ with $\hat h=300$ and
$\alpha=1$, where the polaron is the ground state.

\begin{figure}[t]
	 \centering
	 \includegraphics[width=.4\textwidth]{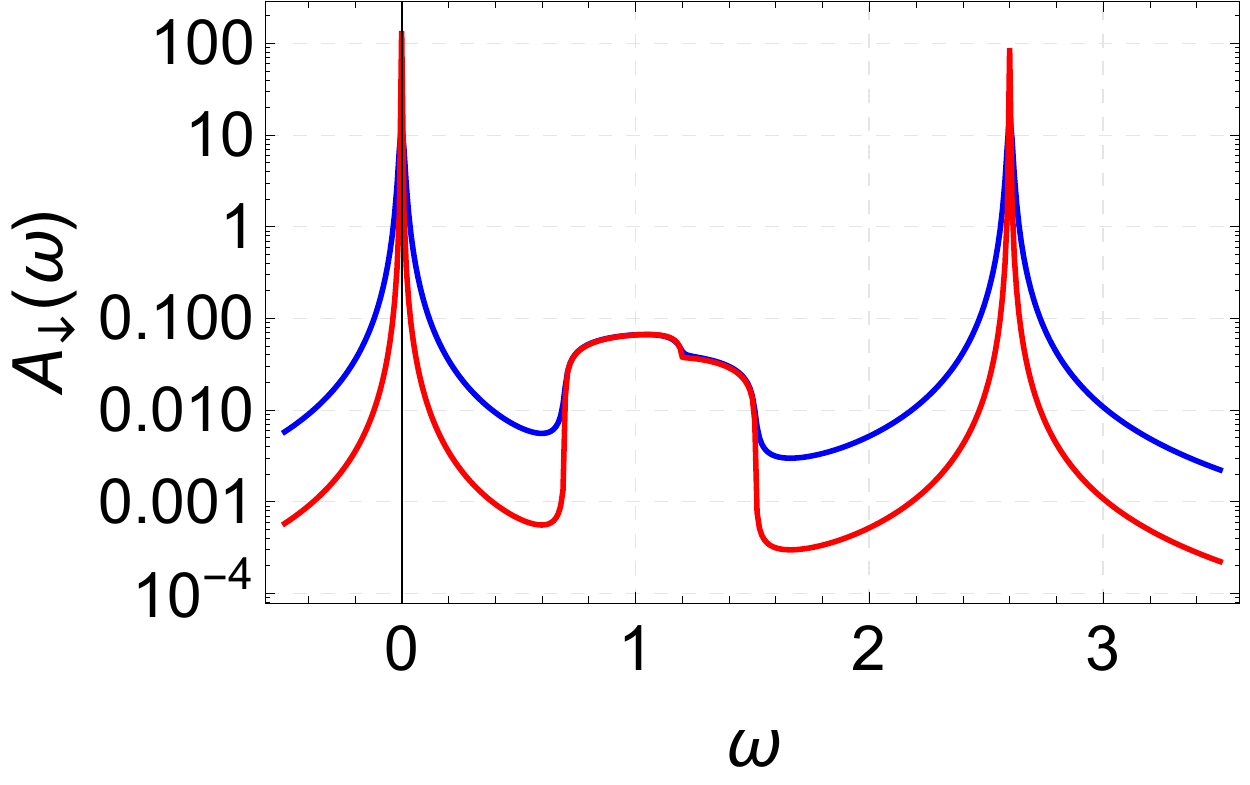}
	 \vspace{-.5\baselineskip}
	 \caption{\label{fg:ed} 
	 Spectral function $\A_{\down}(\hat\omega,\mathbf{0})$ 
	 of an impurity at $1/(k_Fa)=0.7$. The blue (upper) and red
         (lower) curves
	 correspond to $\hat\epsilon=0.01$ and 
	 $\hat\epsilon=0.001$, respectively.}
\end{figure}

Figure \ref{fg:ed} shows the spectral function of a mobile impurity at a 
vanishing momentum.  The two sharp peaks in low and high frequencies
correspond to the attractive and repulsive polarons, respectively.  
The widths of the peaks become narrower for decreasing $\epsilon$, indicating 
that their lifetimes are infinite in our scheme.  On the other hand, in between the attractive and
repulsive polarons, there exists the molecule-hole continuum.  This continuum spectrum 
is stable against a variation of $\hat\epsilon$.

\begin{figure}[t]
	 \centering
	 \includegraphics[width=.4\textwidth]{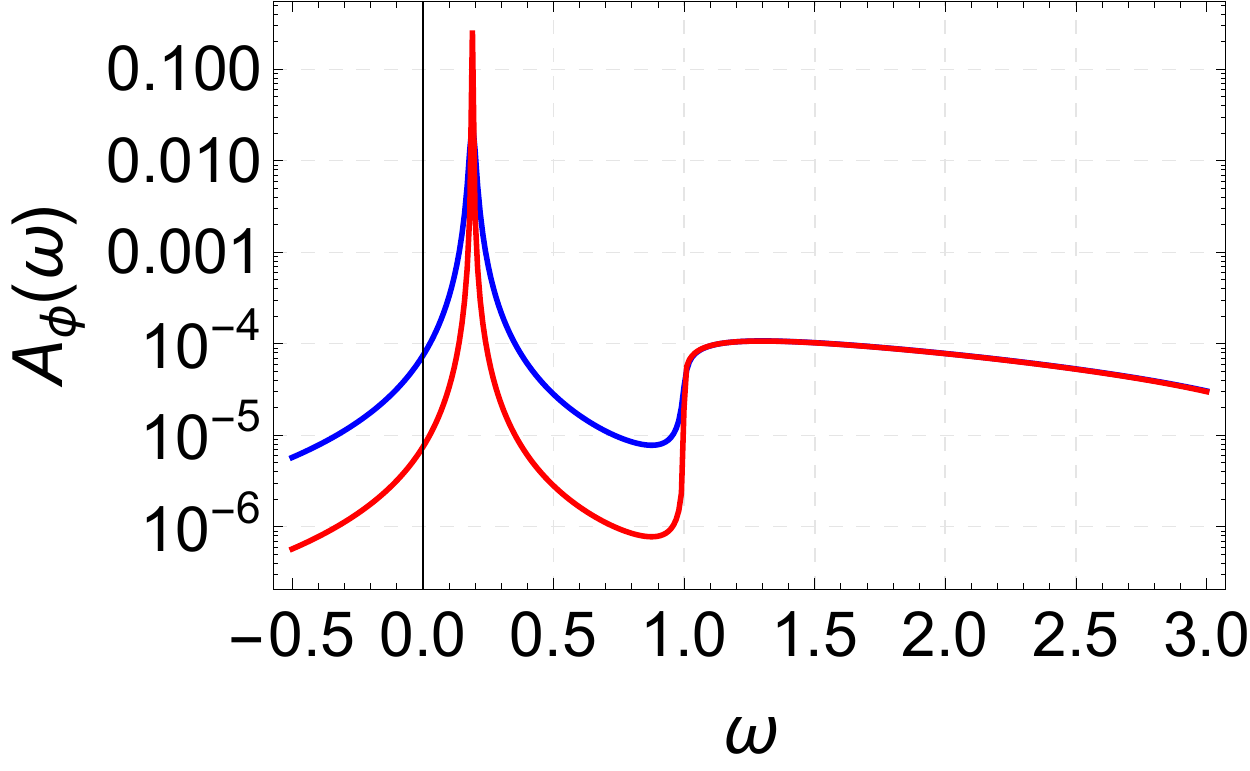}
	 \vspace{-.5\baselineskip}
	 \caption{\label{fg:ep} 
	 Spectral function $\A_{\phi}(\hat\omega,\mathbf{0})$ 
	 of a molecule at $1/(k_Fa)=0.7$. The blue (upper) and red (lower)
         curves
	 correspond to $\hat\epsilon=0.01$ and 
	 $\hat\epsilon=0.001$, respectively.
	 }
\end{figure}

Figure \ref{fg:ep} shows the spectral function of a molecule at a 
vanishing momentum.  There is a single sharp peak at low frequency,
which is the molecule state, and is expected to have an infinite
lifetime.  Similarly to the molecule-hole continuum of the polaron, 
the incoherent background in the high-frequency region 
is robust against a variation of $\hat\epsilon$. 


\section{\label{ap:Tancontact}Tan's contact in a mass-imbalanced Fermi gas}

Here, we discuss the so-called Tan's contact
\cite{Tan20082952,Tan0508,Tan20082987} in the system of 
mass-imbalanced fermions using 
the two-channel model.  Our consideration is based on those in
\cite{Braaten:2008uh,Braaten:2008bi}. Let us 
consider the the following two-channel Hamiltonian:
\ba
	H=&\int_{\mathbf{x}}\bigg[-\sum_{\sigma}\psi_{\sigma}^{*}
	\frac{\nabla^2}{2M_{\sigma}}\psi_{\sigma}
	+\phi^{*}\left(-
	\frac{\nabla^2}{2M_\phi}+\bar{m}^2_{\phi}(a)\right)\phi
	\nonumber
	\\
	&+h(\psi_{\up}^{*}\psi_{\down}^{*}\phi
	+\text{H.c.})
	\bigg],
	\label{eq:two-ch}
\ea
where $\bar{m}^2_{\phi}(a)$ is the mass term introduced in \eqref{eq:mmmf}.
Then, Tan's energy relation \cite{Tan20082952} can be immediately obtained by
considering the expectation value of the above Hamiltonian, which indeed
corresponds to the one discussed in \cite{Braaten:2008uh,Braaten:2008bi} 
in the limit $M_{\up}=M_{\down}$.

Next, we derive the so-called
adiabatic relation \cite{Tan0508}, which is important for our
problem and enables us to calculate Tan's contact directly.  To
this end, let us recall the Feynman-Hellman theorem:
\ba
	\frac{\dd E}{\dd\lambda}=
	\Big\langle\frac{\dd H}{\dd\lambda}\Big\rangle,
\ea
where $\lambda$ is a parameter of the Hamiltonian.        
By setting $\lambda=a$ and using \eqref{eq:two-ch} we obtain
the adiabatic relation 
\ba
	\frac{\dd E}{\dd a}=\frac{C}{8\pi M_r a^2},
\ea
where
\ba
	C=4M_r^2h^2\int_{\mathbf{x}}\langle
	\phi^{*}\phi\rangle
\ea
is nothing but Tan's contact.  Notice that the above expressions
reduce to those in \cite{Braaten:2008uh,Braaten:2008bi} if we set 
$M_{\up}=M_{\down}$.

\bibliographystyle{apsrev4-1}

%

\end{document}